\documentclass[twocolumn,pre,aps
 amsmath,amssymb]{revtex4}

\usepackage{graphics}
\usepackage{epsfig}
\usepackage{dcolumn}
\usepackage{bm}
\usepackage{color}

\begin{document}
\title{Landau-deGennes Theory of Biaxial Nematics Re-examined}
\author{David Allender}
\affiliation{Department of Physics and Liquid Crystal Institute,
Kent State University, POB 5190, Kent OH 44242-0001 USA}
\email{allender@physics.kent.edu}
\author{Lech Longa}
\affiliation{ Jagellonian University, Marian Smoluchowski Institute
of Physics,\\ Department of Statistical Physics and
 Mark Kac Complex Systems Research Center, \\ Reymonta 4, Krak\'{o}w, Poland}
\email{longa@th.if.uj.edu.pl}
\date{\today}
\begin{abstract}
Recent experiments report that the long looked for thermotropic biaxial nematic phase has been
finally detected in some  thermotropic liquid crystalline systems. Inspired by these experimental
observations we concentrate on some elementary theoretical issues concerned with the
classical sixth-order Landau-deGennes free energy expansion in terms of the symmetric and traceless tensor order parameter $Q_{\alpha\beta}$. In particular, we fully explore the stability of  the biaxial nematic phase giving analytical solutions for all distinct classes of the phase diagrams that theory allows. This includes diagrams with triple- and (tri-)critical points and with multiple (reentrant) biaxial- and uniaxial phase transitions. A brief comparison with predictions of existing molecular theories is also given.
\end{abstract}
\maketitle

\section{INTRODUCTION} \label{sec:Introduction}
The biaxial nematic phase, predicted theoretically by Freiser \cite{Freiser:prl:biax:1970,Freiser:mclc:biax:1971} over 35 years ago,  is one of the perennially challenging problems of experimental soft-matter physics. Although discovery of this phase was made by Saupe and co-workers in a fine-tuned lyotropic liquid crystal system in 1980 \cite{Yu:prl:biax:1980}
only in the past three years- and following several earlier attempts that proved unsuccessful in this regard (\emph{for a comprehensive review
see e.g.} \cite{Luckhurst:thinSolFilms:biax:rev:2001,%
Luckhurst:nature:biax:rev:2004})- strong experimental evidence has become available that this phase can also be made stable in
thermotropic  liquid crystalline materials  \cite{Madsen:prl:biax:2004,Acharya:prl:biax:2004,%
Merkel:prl:biax:2004,Madsen:prl:biax:replyToGalerne2006%
}. This discovery raises the emerging  theoretical problem of what mechanism is responsible for the stability of thermotropic biaxial nematics, especially for bent-core systems \cite{Madsen:prl:biax:2004,Acharya:prl:biax:2004} and for tetrapode-like molecules \cite{Merkel:prl:biax:2004,Neupane:prl:lightScatteringTetrapodes:2006}, where this phase was shown to be stable.

There are two nematic phases of distinct symmetries. The ubiquitous uniaxial nematic phase has the ${\cal{D}}_{\infty h}$ point group symmetry \cite{DeGennes:book,Gramsbergen:pr:nem:1986,%
Singh:review:2000,Longa:pre:biax:1994}, which  results in the definition of a single mesoscopic direction, known as the \emph{director}. The director is a unit   vector,  denoted {$\mathbf{\hat{n}}$}, with
the  directions  {$\mathbf{\bf \hat{n}}$}  and  {$
-\mathbf{\hat{n}}$} being equivalent.
One consequence of this is that there are two
different principal components of a second rank tensorial property, such as {\emph{e.g}} the magnetic susceptibility.
Generally, two uniaxial
nematic phases are distinguished: prolate ($N_{U+}$) and oblate ($N_{U-}$). The prolate uniaxial states  usually occur for rod-like molecules while disc-like molecules
yield the oblate uniaxial states. As opposed to the uniaxial nematic phase, the biaxial nematic phase, denoted  $~N_{B}$, is characterized by  three orthonormal directors, the Goldstone modes, which we denote $\{
\mathbf{\hat{l}}, \mathbf{\hat{m}}, \mathbf{\hat{n}}=
\mathbf{\hat{l}} \times \mathbf{\hat{m}}\} $.
Due to overall lack of polarity of the known biaxial nematics,
one finds that $\mathbf{\hat{l}}$ and  $-\mathbf{\hat{l}}$,
  $\mathbf{\hat{m}}$   and  $-\mathbf{\hat{m}}$  and
  $\mathbf{\hat{n}}$   and  $-\mathbf{\hat{n}}$
directions are equivalent. That is, from the symmetry point of view the biaxial nematic phase is  a structure of
${\cal{D}}_{2 h}$ point-group symmetry and the corresponding second rank tensorial property has three
different principal components.

Generally, first-- and second   order phase   transitions   are
observed experimentally   between the isotropic phase and different nematic phases and between the nematic phases.
The phase sequence of $Iso$ $ \leftrightarrow $ $(N_{U-})$
$\leftrightarrow$ $(N_{B})$ $\leftrightarrow$ $(N_{U+})$
$\leftrightarrow$ $(N_{B})$ $\leftrightarrow$ $(N_{U-})$
$\leftrightarrow$ $(Iso)$ is found with decreasing temperature
\cite{Yu:prl:biax:1980,Charvolin:nc:lyotr:biax:1984,%
Figueiredo:book:lyotr:2005,Merkel:prl:biax:2004}, where the brackets indicate that some of the phases may not appear. In particular, the amazing  reentrant uniaxial and isotropic phases are observed in lyotropic systems (\emph{see e.g.}
 \cite{Figueiredo:book:lyotr:2005} {and references therein}).

On the theoretical level, possible effects of molecular structure on nematic
order have been studied. More specifically, molecular
field theories of single-component systems consisting of biaxial
molecules and interacting via hard-core or continuous potentials were shown to produce a stable biaxial phase
\cite{Freiser:prl:biax:1970,Alben:prl:biax:1973,%
Mulder:pra:biaxBif:1989,Teixeira:mclc:biax:1998,%
Sonnet:pre:biax:2003,Longa:pre:biax:2005,Longa:pre:biaxSU3:2007}.
A similar scenario emerges  from computer simulation studies
\cite{Biscarini:prl:biax:1995,Cleaver:pre:biax:1996,Sarman:jcp:biax:1996,%
Camp:jcp:biax:1997,Ginzburg:cp:simBiax:1997,Camp:jcp:biax:1999,%
Berardi:jcp:biax:2000}
and from Landau treatments
\cite{Allender:mclc:biax:1984,Allender:mclc:biax:1985,%
Gramsbergen:pr:nem:1986,Prostakov:cr:biaxpheno:2002}.

Out of the theories cited the simplest description of the uniaxial and biaxial nematic phases is one offered by a sixth-order Landau-deGennes free energy expansion in terms of the alignment tensor $Q_{\alpha\beta}$ (\ref{eq:degl0}). The  theory  is
generally employed to interpret experimental data as well as to classify possible topologies of the phase diagrams. Therefore it seems quite important to know, if possible,  an analytical form of all distinct classes of the phase diagrams and limitations on them that can be derived from this simple theory.
This  task has only partly been realized so far \cite{Allender:mclc:biax:1984,Allender:mclc:biax:1985,%
Gramsbergen:pr:nem:1986,Toledano:pre:biaxdiag:1995,Longa:revBras:1998,%
Singh:review:2000,Prostakov:cr:biaxpheno:2002}.
None of the papers cited shows, however, a full spectrum of predictions of this theory.
The closest to the ideal is the paper by Prostakov \cite{Prostakov:cr:biaxpheno:2002}, but also there not all cases/analytical solutions have been given.

Owing to the current excitement in the field of thermotropic biaxial nematics  we think it is important to re-examine this fundamental theory. We give analytical formulas for all distinct classes of the phase diagrams the model can predict and for their stability range. We hope this will be of some help for experimentalists in analyzing experimental data on biaxial nematics and
will bring partial order to existing molecular predictions on this phase.

This paper is organized as follows. After a brief  discussion
of the Landau-deGennes theory in Sec.~II, we
give analytical solutions for the phase diagrams in Sec.~III.
Section IV is  devoted  to a short discussion.
\section{Landau-deGennes Free Energy}
The  best way to account for a symmetry change that takes place across a phase transition  is by referring to an order parameter. For a phenomenological description of the nematics the relevant order parameters are tensors built out of the directors. Among these the leading order parameter is the second rank symmetric and traceless alignment tensor $ \mathbf{Q}$. In a standard
parametrization $\mathbf{Q} $ can be written as
\begin{eqnarray}\label{eq:Q-diagonal}
\mathbf{Q} &=& {\frac{q_0}{\sqrt{6}}}\left(3\hat{\mathbf{n}}\otimes \hat{\mathbf{n}} -
{\boldsymbol{1}} \right) +
{\frac{q_2}{\sqrt{2}}}\left(\hat{\mathbf{l}}
\otimes \hat{\mathbf{l}} - \hat{\mathbf{m}}
\otimes \hat{\mathbf{m}}\right),
\end{eqnarray}
where  the
directors \{$\mathbf{\hat{l}},\mathbf{\hat{m}}, \mathbf{\hat{n}}$\}
are identified with  eigenvectors   of $\mathbf{Q}$ corresponding to the eigenvalues
$\lambda_1=-\frac{{q_0}}{\sqrt{6}}+\frac{{q_2}}{\sqrt{2}}$,
$ \lambda_2=-\frac{q_0}{\sqrt{6}}-\frac{q_2}{\sqrt{2}}
$,
$ \lambda_3=-\lambda_1-\lambda_2=\sqrt{\frac{2}{3}} q_0
 $,
respectively. The parametrization (\ref{eq:Q-diagonal}) for $\mathbf{Q}$ is chosen such that the formula for $F$ is kept concise. The   isotropic state is stabilized  when all three eigenvalues of
${\mathbf{Q}} $ are equal and hence vanish, which  yields
${\mathbf{Q}}\equiv {\mathbf{0}}$. For the ${\cal{D}}_{\infty h}$-
symmetric uniaxial states two out of the three eigenvalues of
${\mathbf{Q}}$ are equal, {\em i.e.}, $q_0 \neq 0,\, q_2=0$ or
$q_0 \neq 0,\, q_2=\sqrt{3}\,q_0$ or  $q_0 \neq 0,\, q_2=-\sqrt{3}\,q_0$. In  the general case, ${\mathbf{Q}}$ has three different real eigenvalues that account for the
${\cal{D}}_{2 h}$- symmetric biaxial state. A microscopic interpretation of the alignment tensor for simple molecular models is found in \cite{Mulder:pra:biaxBif:1989,Longa:lc:Qtensor:MolInterpretation2005}
and can easily be extended to the more general cases.

The Landau-deGennes phenomenological theory of non-chiral systems is implicitly based on the hypothesis that equilibrium  properties of the system can be found from a non-equilibrium free energy, constructed as an {${\cal{O}}(3)$}--symmetric expansion in powers of $\mathbf{~Q}$.
The only restriction on the expansion is that it must be stable
against an unlimited growth of the order parameter. There are two
types of ${\cal{O}}(3)$ invariants that can be constructed out of $\mathbf{Q}$, which involve traces and determinants of powers of $\mathbf{Q}$. But determinants can be expressed
in terms of traces and all traces of
$\mathbf{Q}^n$ with $n \ge 4$ are polynomials of
$\mathrm{Tr}{(\mathbf{Q}}^{2})$ and $\mathrm{Tr}{(\mathbf{Q}}^{3})$ \cite{Gramsbergen:pr:nem:1986}. In addition, $\mathrm{Tr}{(\mathbf{Q}}^{2})$ and
$\mathrm{Tr}{(\mathbf{Q}}^{3})$ are bounded by the inequality
\begin{eqnarray}\label{eq:traceRestriction}
    \frac{1}{6} \mathrm{Tr}{(\mathbf{Q}}^{2}) ^3
     - \mathrm{Tr}{(\mathbf{Q}}^{3}) ^2 =\hspace{4cm}\nonumber\\
      \hspace{-2cm} \frac{1}{3}
   \left(\lambda _1-\lambda _2\right)^2
   \left(2 \lambda _1+\lambda _2\right)^2
   \left(\lambda _1+2 \lambda _2\right)^2 \ge 0,
\end{eqnarray}
which is fulfilled as equality for the uniaxial phases.

\begin{figure}
\includegraphics*[width=8cm,height=5cm]{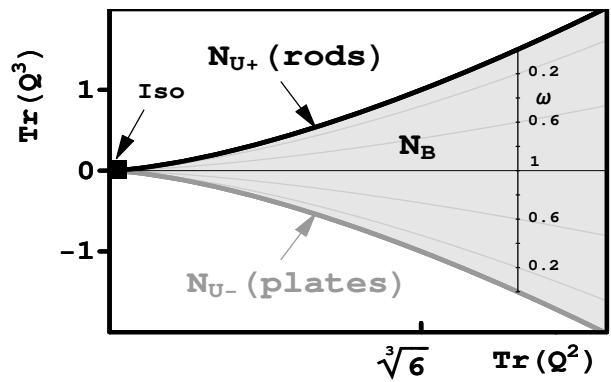}
\caption{Allowed variation of the independent degrees of freedom
 $\mathrm{Tr}{(\mathbf{Q}}^{2})$ and $\mathrm{Tr}{(\mathbf{Q}}^{3})$
 (shaded area) and identification
 of the corresponding phases. Shown are also lines of constant
 biaxiality parameter, $\omega$.
\label{fig:trq3Vstrq2}}
\end{figure}

A coordinate-independent form of the inequality
(\ref{eq:traceRestriction}) is obtained by a very convenient
re-parametrization in $\mathrm{Tr}(\mathbf{Q}^{2})$
and $\mathrm{Tr}(\mathbf{Q}^{3})$ that uses just two scalar
parameters: $q$ and $0 \le \omega \le 1$. They are  introduced
through the relations
\begin{eqnarray}
\mathrm{Tr}(\mathbf{Q}^{2})&=& q^2=q_0^2+q_2^2 = I_2
\label{eq:uniaxiality1}
\\
{\sqrt{6}}\,\mathrm{Tr}(\mathbf{Q}^{3})   &=& {q^3(1-\omega)} =
q_0^3 -3q_0q_2^2=I_3, \label{eq:uniaxiality2}
\end{eqnarray}
where $|q|$ is the norm of $\mathbf{Q}$
and  $\omega$ serves as a normalized measure of phase biaxiality.
The ${\cal{D}}_{2 h}$- symmetric biaxial state is characterized by
$\omega >0$ with maximal biaxiality being accomplished  for
$\omega=1$. For the uniaxial phases $\omega = 0$. In addition, for uniaxial $\mathbf{Q}$-tensors a transformation
$\mathbf{\hat{u}}\rightarrow [(Q_{\alpha\beta} -
c\delta_{\alpha\beta})u_{\alpha}u_{\beta}]$, where c is an arbitrary
constant making the bilinear form $[...]$ positive-definite,
transforms  a unit sphere $|\mathbf{\hat{u}}|=1$ into an axially
symmetric, prolate- ($q>0$) or oblate ($q<0$) closed surface. Hence the sign of $q$, being consistent with the sign of
$\mathrm{Tr}(\mathbf{Q}^{3})$, allows one to distinguish between
${N_{U+}}$ ($q>0$) and ${N_{U-}}$ ($q<0$) phases. Actually $q$ and $\omega$
can serve as invariant measures of order in
uniaxial ($q\ne0, \omega=0$)- and biaxial ($q\ne0, \omega\ne0$)
nematics. For the isotropic phase $q=0$.
The allowed variation of  $\mathrm{Tr}(\mathbf{Q}^{2})$
and $\mathrm{Tr}(\mathbf{Q}^{3})$ and consequently also of $q$ and $\omega$, along with the identification of
different nematic phases, is shown in Fig.~(\ref{fig:trq3Vstrq2}).

In the absence of electric and magnetic fields the bulk free energy for the isotropic- and the nematic phases has the form
\begin{equation}\label{eq:F}
    F(\mathbf{Q})= F[\mathrm{Tr}{(\mathbf{Q}}^{2}),
    \mathrm{Tr}{(\mathbf{Q}}^{3})]=F[I_2,I_3]=F[q,\omega].
\end{equation}
The\emph{ minimal coupling Landau expansion} of $F$ that accounts for the biaxial nematic phase has to be taken up to 6th order with respect to $\mathbf{Q}$. This theory, also known as Landau-deGennes free energy of biaxial nematics, reads ({\em see e.g. }
\cite{Gramsbergen:pr:nem:1986})
\begin{eqnarray} \label{eq:degl0}
 \hspace{-2mm} F &=& F_{o} +
   \frac{1}{2}a\, I_2 -
  \frac{1}{3}b\, I_3 +
  \frac{1}{4}c\,I_2^{2} +
           \frac{1}{5}d \, I_2 I_3
            +   \nonumber\\
   && \frac{1}{6}e \,I_2^{3} +\frac{1}{6}(f- e)
                I_3^{2} + \ldots \nonumber \\
                & =& F_{o} + F_u(q) + F_b(q)\omega  +
                \frac{1}{6}\, \phi\, q^6 \omega ^2
                \ldots
\end{eqnarray}
with
\begin{eqnarray}
%
  F_u &=& \frac{1}{2}a \,q^2-\frac{1}{3}b \,q^3+\frac{1}{4}c
  \, q^4+\frac{1}{5}d\, q^5+\frac{1}{6}f\, q^6 \hspace{1cm}
  \label{eq:degl0-definitions1}\\
  F_b &=& \frac{1}{3}b\, q^3-\frac{1}{5}d\,
   q^5 -\frac{1}{3}\,\phi\, q^6.
   \label{eq:degl0-definitions2}
\end{eqnarray}
In Eq.~(\ref{eq:degl0}) the $F_{o}$-part represents the unimportant  free energy of the reference isotropic
phase; $F_u$ is the free energy of the uniaxial phases ($\omega=0$)
and the remaining two terms represent biaxial contributions. The
coefficients of the expansion generally depend on temperature (inverse density) and other thermodynamic (control) parameters. In what follows we will only keep an explicit dependence on the temperature. In particular, the
coefficient $a=a_o( T - T^{*} )$ with $T$ being the absolute
temperature, is the only term in the expansion that is assumed to be explicitly temperature (or density)-dependent. On general
thermodynamic grounds (see {\em e.g.} \cite{Longa:lc:Qtensor:MolInterpretation2005}) one can show  that $a$ is usually the first of the coefficients in the expansion (\ref{eq:degl0}) that changes sign  as temperature is lowered. The sign change is a result of competition between either energy and entropy or different forms of entropy. The parameter $T^{*}$ accounts quantitatively for this competition and represents the spinodal temperature for the first order phase transition from the isotropic phase to the uniaxial nematic phase. As for the remaining parameters: $a_o >0$ by definition and stability of the expansion requires $e>0$ and $f>0$. Except for cases of  multicritical behavior, the signs of $b,c,d,e,f$ are assumed not to change in the vicinity of $T^{*}$. Hence, these coefficients,  being weakly temperature-dependent, are assumed constants and taken at  $T=T^{*}$.

The parametrization of $F$ in terms of $q$ and $\omega$, Eq.~(\ref{eq:degl0}), leads to a simple determination of absolute minima of $F$ and, hence, a
construction of the corresponding phase diagrams. Clearly, the form of $ \mathbf{Q}$, Eq.~(\ref{eq:Q-diagonal}), implies that the
\emph{Iso}-$N_B$ and the $N_U-N_B$ phase transitions can be either
first- or second order. In other words we may expect first-order,
second-order and  tricritical behavior  at \emph{Iso}-$N_B$ and
$N_U-N_B$ transitions, depending on model parameters.
\section{Phase diagrams}
 Out of the five material parameters
$b,c,d,e,f$ ($\phi=f-e $), introduced in Eq.~(\ref{eq:degl0}), two
are redundant and can be set equal to 0 or $\pm 1$. This is a direct consequence of the freedom to choose a scale for the free energy and for ${\mathbf{Q}}$. If not specified otherwise we choose  $e=1$ and $c=0,\pm1$, and investigate the phase diagrams in the ($a$, $b$)-plane as function of  $d$ and $f$.
Additionally,  we assume  $f>0$ to guarantee the stability of the
expansion (\ref{eq:degl0}) against unlimited growth of $q$ and
replace $f-e$ by $\phi$ ($\phi=f-e\equiv f-1$)  whenever convenient. We also make use  of the
free energy invariance with respect to the transformation:
$\{b,d,q\} \rightarrow \{-b, -d, -q\}$, which limits $d$ to $d \ge
0$. The diagrams for $d<0$ are obtained as mirror images with
respect to the $b=0$ line of those for $d>0$, followed by  a
 subsequent change of $N_{U\pm}$ into $N_{U\mp}$.

Interestingly, the relatively simple expansion (\ref{eq:degl0})
generates a rich spectrum of possibilities for phase
diagrams. We
show that all of them can be  divided  into ten distinct classes,
where four  involve only uniaxial phases. The remaining cases,
corresponding to  $d < 0$, are obtained from the classes discussed
by applying the aforementioned $b=0$ mirror transformation.
\subsection{Phase diagrams with uniaxial phases: $q\ne 0,\omega = 0$}
We start by considering regions of stability of the uniaxial
nematic.  The necessary conditions for this phase
to become, at least, locally stable read
\begin{eqnarray}\label{eq:uniaxial-necessary}
%
  \frac{\partial F_u}{\partial q} &=& q \left( a- b q +c q^2+ d
   q^3  + f q^4 \right) =0 \\ \label{eq:uniaxial-stability}
  \frac{\partial^2 F_u}{\partial q^2} &=& a- 2 b q +
  3 c q^2+ 4 d q^3 + 5 f q^4 > 0.
\end{eqnarray}
The limit of local stability is attained when  the inequality
(\ref{eq:uniaxial-stability}) becomes   equality, which, together
with (\ref{eq:uniaxial-necessary}), describes  a  saddle bifurcation
in the model and  represents spinodal lines.
These conditions are particularly simple to solve for $a$ and $b$ in a parametric, q-dependent form.  The nontrivial solution is
\begin{eqnarray}\label{eq:ab-uniaxial-parametric}
  a &=& c q^2 + 2 d q^3 + 3 f q^4\\
  b &=& 2 c q +  3 d q^2  + 4 f q^3,
\end{eqnarray}
which, together with the trivial one: $\left\{a=0,q=0,(\forall\,
b)\right\}$  defines the borders of the area in the $(a,b)$ plane,
where the solutions to the Eq.~(\ref{eq:uniaxial-necessary})
are, at least, locally stable. Clearly, $q$ runs over all real numbers. The subsequent  calculation of the
free energy at these local minima allows us to select the global
minimum within the family  of uniaxial solutions.

A complete analysis of the model, including calculation of the free
energy, proceeds in a similar way. In particular, we determine
parametrically the transition line between the isotropic- and the
uniaxial phases by solving the system of equations: $\{F_u=0, \partial
F_u/\partial q=0\}$ for $a(q)$ and $b(q)$. The solution reads
\begin{eqnarray}\label{eq:ab-uniaxial-Isotropic-parametric}
  a &=& \frac{c q^2}{2} +\frac{4 d q^3}{5} + f q^4   \\
  b &=& \frac{3 c q}{2}+\frac{9 d q^2}{5} + 2 f q^3.
\end{eqnarray}
Subsequent analysis of the Eqs.~%
(\ref{eq:ab-uniaxial-parametric},
\ref{eq:ab-uniaxial-Isotropic-parametric})
allows us to single out  four topologically distinct classes of the
phase diagrams with uniaxial- and isotropic phases. The
representatives of each class are shown in
Figs.~\ref{fig:nu-iso}-\ref{fig:nu-nu-nu-iso-degenerate}.
The
corresponding global stability sectors in the $\{c,d,f\}$ parameter
space are given in
Figs.\ref{fig:df-for-c-eq-1}-\ref{fig:df-for-c-eq-m1}.

We shall now characterize each of the 'uniaxial' classes of the
diagrams.
\subsubsection{Class (a)}
\begin{figure}
\includegraphics*[width=8cm,height=8cm]{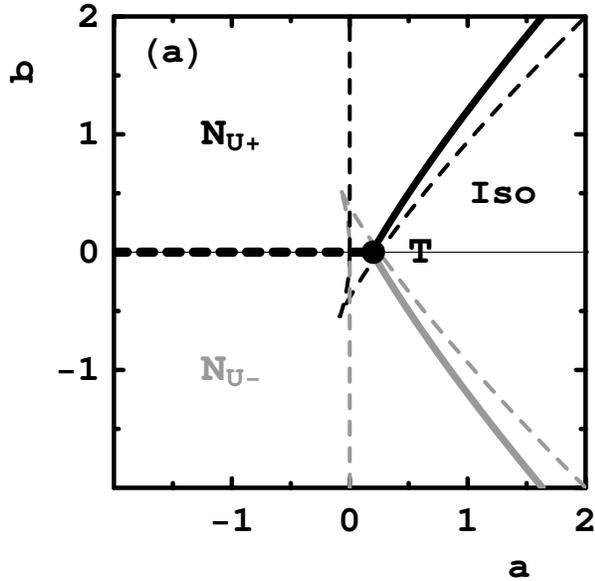}
\caption{Generic phase diagram with a direct, first order phase
transition from isotropic- to uniaxial nematic phases, where gray
color refers to  phase transitions involving the  $N_{U-}$ phase.
For $b=0$ a phase transition between  isotropic and highly
degenerated phase of $q\ne0$ but of arbitrary $\omega$ takes place
at the quadruple point (T). For $c>0$ T became a tetracritical
point. Parameters taken are $(c,d,f)=(-1,0,1)$. Thin dashed lines,
representing spinodal, are the solutions of the
Eq.~(\ref{eq:ab-uniaxial-parametric}); also $a=0$ spinodal is shown.
\label{fig:nu-iso}}
\end{figure}

The first  class  is obtained for $d=0$, Fig.~\ref{fig:nu-iso}.
It contains  a line of  first order $Iso\leftrightarrow{N_{U+}}$ phase transitions for $b > 0$, a line of first order
$Iso\leftrightarrow{N_{U-}}$ phase transitions for $b < 0$, and a degenerated biaxial phase of $q\ne0$ and arbitrary $\omega$, stable
only along the $b=0$ line. We shall come back to the degenerated case in the last subsection of this paper. The $Iso\leftrightarrow{N_{U\pm}}$ lines
have a common tangent  $a=0$. For $c\ge 0$ the four lines meet at an
isolated, tetracritical point, also often referred to as the Landau
point. Its coordinates are $(a,b)=(0,0)$. For $c<0$ the Landau point
becomes a quadruple point of coordinates $(a,b)=(\frac{3 c^2}{16 f}
,0)$, marked  as 'T' in Fig.~\ref{fig:nu-iso}.  The whole phase
diagram is given analytically by
\begin{equation}\label{eq:isoNuFord_0}
    b^2=\frac{\sqrt{\left(c^2+16 a f\right)^3}-c \left(c^2-48 a
f\right)}{16 f},
\end{equation}
where  $a\geq \frac{3 c^2}{16 f}\Theta(c)$ with $\Theta$ being the
step function.

Now we concentrate on  more  complex cases with $d\ne0$. They are
gathered in three classes of the diagrams, denoted (b)-(d).
\subsubsection{Class (b)}
\begin{figure}
\includegraphics*[width=8cm,height=8cm]{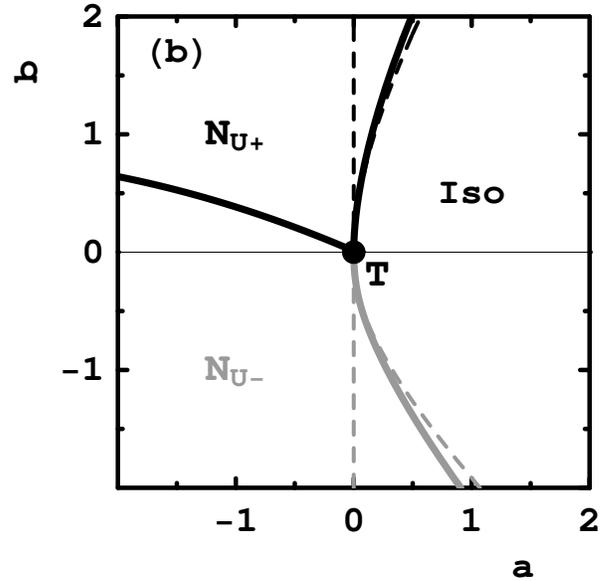}
\caption{Generic phase diagram with a direct, $N_{U+}-N_{U-}$ first
order phase transition. Parameters taken are $c=d=f=1$. For detailed
definition of all symbols and lines used see caption to Fig.~\ref{fig:nu-iso}.
The triple point (T) is localized at $(a,b)=(0,0)$.
\label{fig:nup-num-iso}}
\end{figure}

Diagrams that belong  to this class are similar to (a)
except for  an additional first-order phase transition line between
the ${N_{U+}}$ and ${N_{U-}}$ phases. A typical situation is shown
in Fig.~\ref{fig:nup-num-iso}.
Also in this case  the  transition
line can be  given in an analytical form as:
\begin{eqnarray}\label{eq:NUp-NUm}
    a&=&u \left[ 2 d u (d+2 f u)+c (2 d+5 f u)\right]\times\nonumber\\
        && \frac{    \left[5 c+2 u (4 d+5 f u)\right]}{4 (d+5 f u)^2}\\
    b&=&-\frac{u \left[3 c d+2 u
         \left(3 d^2+7 f u d+5 f^2 u^2\right)\right]}{2 (d+5 f u)}
\end{eqnarray}
with the free energy
\begin{equation}\label{eq:NUp-NUm-Fu}
    F_u= -\frac{u^2 (3 d+5 f u) \left[5 c+2 u (4 d+5 f u)
    \right]^3}{240 (d+5 f
    u)^3}.
\end{equation}
The parameter u  must satisfy the inequalities
\begin{eqnarray}
    -\frac{3 d}{5 f}&\leq u\leq& -\frac{d}{5 f}
    \hspace{5mm}\text{for}\hspace{5mm}
    25 c f\leq 6 d^2 \label{eq:NUp-NUm-parameter-range-a}\\
-\frac{d}{5 f}&\leq u\leq &0 \hspace{10.5mm}\text{for}\hspace{5mm}
25 c f\geq 6 d^2.\label{eq:NUp-NUm-parameter-range-b}
\end{eqnarray}
Generally, this topology is observed for the $\frac{c f}{d^2}$
parameter taken from outside of the interval $[\frac{6 }{25 },
\frac{9 }{25}]$ (\emph{see} discussion below leading to inequalities
(\ref{eq:c-inequality-uniaxial-a}, \ref{eq:c-inequality-uniaxial-b})) and is the most typical for the
uniaxial family of the phase diagrams,
Figs.~\ref{fig:df-for-c-eq-m1}-\ref{fig:df-for-c-eq-1}. The appearance of the
${N_{U+}}- {N_{U-}}$ line is a result of competition between the third- and the fifth order invariants in the free energy expansion when the coefficients weighting these terms are of the opposite
sign. For ${25}c f \ge 9 d^2$ the three phases: $Iso$, $N_{U+}$ and $N_{U-}$ meet at the triple point, T, of coordinates $(a,b)=(0,0)$
where $u=0$ and $F_u=0$. At T the lines $Iso-N_{U+}-$ and
$Iso-N_{U-}$ have a common tangent given by the $a=0$ line. For ${25}cf \le 6 d^2$ the triple point moves away from the origin to a new location at
\begin{eqnarray}\label{eq:NU-NU-Iso-triple-point}
    a&=&\frac{3 \left(6 d^2-25 c f\right)^2}{10000 f^3}\\
    b&=&\frac{9 d \left(6 d^2-25 c f\right)}{500
   f^2},
\end{eqnarray}
which is obtained by substituting $u=-\frac{3d}{5f}$ into
Eqs.~(\ref{eq:NUp-NUm}).
\subsubsection{Class (c)}
This class of the phase diagrams is perhaps the most interesting one
among the uniaxial topologies. In addition to the ${N_{U+}}-
{N_{U-}}$ transition line,   shown in Fig.~\ref{fig:nup-num-iso} and
given by (\ref{eq:NUp-NUm},\ref{eq:NUp-NUm-parameter-range-b}), it
also displays a direct $N_{U-}-N_{U-}$ first-order phase transition
line terminating at a critical point of the liquid-vapor type.

\begin{figure}
\includegraphics*[width=8cm,height=8cm]{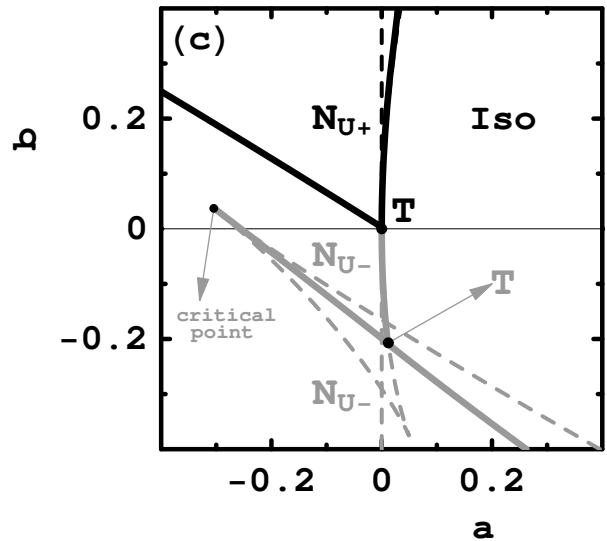}
\caption{Generic phase diagram with a direct,  $N_{U-}-N_{U-}$ first
order phase transition terminating at a critical point. Parameters
taken are $(c,d,f)=(1,1.05,0.306)$. Note that one of the spinodal
lines connected with the $N_{U+}$ phase is separated from the
transition line $Iso-N_{U+}$ by less than the thickness of the graph,
and, hence, is invisible. For detailed definition of all symbols
and lines used see caption to Fig.~\ref{fig:nu-iso}.\label{fig:nu-nu-nu-iso}}
\vspace{2cm}
\end{figure}
Again
this behavior results from the aforementioned competition between
the third- and the fifth order terms in the free energy expansion.
An example of the ${N_{U-}}\leftrightarrow{N_{U-}}$ line, together
with the lines: $~Iso\leftrightarrow{N_{U-}}$,
$~Iso\leftrightarrow{N_{U+}}$ and
$~Iso\leftrightarrow{N_{U+}}\leftrightarrow{N_{U-}}$, is shown in
Fig.~\ref{fig:nu-nu-nu-iso}. The lines terminate at the
$Iso-N_{U+}-N_{U-}$ triple point of coordinates $(a,b)=(0,0)$ and at
the $Iso-N_{U-}-N_{U-}$ triple point given by the formula
(\ref{eq:NU-NU-Iso-triple-point}). The necessary condition for this
class of the diagrams to appear is a requirement that the spinodal
has two cuspidal points for the oblate states ($q<0$). After
inspecting  $q$-dependence of the curve
(\ref{eq:ab-uniaxial-parametric}) one easily finds that the
$(a,b)$-coordinates  of these points are obtained by substituting
\begin{equation}\label{eq:q-cusps}
  q_\pm =  \frac{-d \pm \sqrt{d^2-\frac{8 }{3}c f}}{4 f}
\end{equation}
into  Eq.~(\ref{eq:ab-uniaxial-parametric}). Additionally, the
conditions $0 \le { 8 c f} < {3} d^2\,\,\,\, \wedge \,\,\,\, c>0 $
(in our re-scaling $c=1$) must  be met for the cuspidal points with
negative values of $q$ to occur. Taken together, these conditions
guarantee that there exist two local minima (usually one of them
becomes the global one) and two local maxima in the free energy
branch for the oblate states ($q<0$). The local minima can finally
convert into a stable $N_{U-}-N_{U-}$ line, Eq.~(\ref{eq:NUp-NUm}),
if
\begin{eqnarray}
    {6 }{}d^2\le 25 {cf}{}
\le {9 }{}d^2\,\,\,\,\,  &\wedge& \,\,\,\,\, c>0
\label{eq:c-inequality-uniaxial-a}\\
-\frac{d}{2 f}-\frac{\sqrt{9 d^2-24 c f}}{6 f}\leq & u&\leq -\frac{3
d}{5 f}.\label{eq:c-inequality-uniaxial-b}
\end{eqnarray}
Note that the conditions (\ref{eq:c-inequality-uniaxial-a},\ref{eq:c-inequality-uniaxial-b}) are more
restrictive than the ones for the cuspidal points of the spinodal.
The first one, (\ref{eq:c-inequality-uniaxial-a}), states that the
$Iso-N_{U-}-N_{U-}$ triple point disappears (and hence also the
$N_{U-}-N_{U-}$ line) for $b>0$. Additionally,  it  guarantees the
appearance of the $Iso-N_{U-}-N_{U-}$ triple point (cusp) in the
$F_u(q<0)=0$ branch of the free energy for $b<0$. The second
inequality, (\ref{eq:c-inequality-uniaxial-b}), represents actually
the same restrictions, but expressed in terms of u. Finally,
coordinates of the critical point are obtained by substituting
$q_{-}$, Eq.~(\ref{eq:q-cusps}), into
Eq.~(\ref{eq:ab-uniaxial-parametric}). This leads to
\begin{eqnarray}\label{eq:NU-NU-Critical-Point}
  a &=&
  \frac{4 c f \left(3 d^2-2 c f\right) - 3 d^4}{96 f^3}
  -\frac{d \left(3 d^2-8 c f\right)^{3/2}}{96 \sqrt{3} f^3}\hspace{3mm}\\
  b &=&\frac{9
   d^3-36 c f d+\sqrt{3} \left(3 d^2-8 c f\right)^{3/2}}{72 f^2}.
\end{eqnarray}
Sector of stability of the class (c) is shown in
Fig.~\ref{fig:df-for-c-eq-1}. It is restricted to the area given by
$6d^2/25<f<9d^2/25 \wedge f<1$. The richest phase sequence obtained for this class as temperature is lowered is $Iso-N_{U+}-N_{U-}-N_{U-}$.
\subsubsection{Class (d)}
\begin{figure}
\includegraphics*[width=8cm,height=8cm]{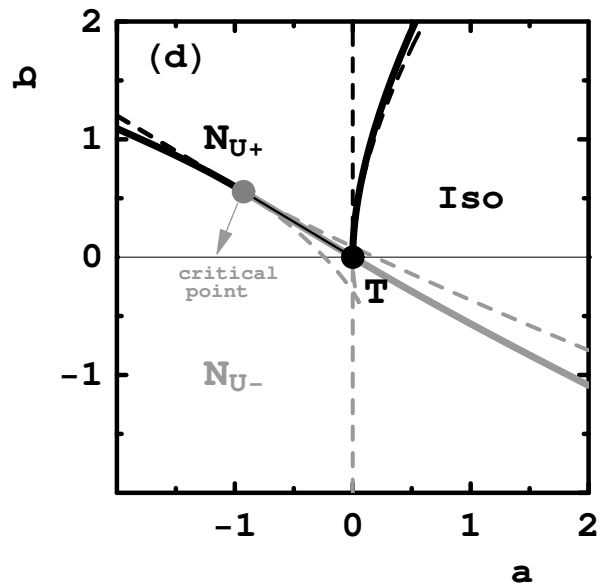}
\caption{Degenerated version of  phase diagram shown in
Fig.~\ref{fig:nu-nu-nu-iso}. The $N_{U-}-N_{U-}$ line (together with
the critical point) belongs to the $N_{U-}-N_{U+}$ line starting at
$a=0$, both being straight lines. The dashed straight line above the
critical point is a continuation of the $N_{U-}-N_{U-}$ line and
serves as a reference to $N_{U-}-N_{U+}$ line. Parameters taken are
$(c,d,f)=(1,1,6/25)$. For detailed definition of all symbols and lines used
see caption to
Fig.~\ref{fig:nu-iso}.\label{fig:nu-nu-nu-iso-degenerate}}
\end{figure}

Quite interesting and  untypical situation is met when  $cf/d^2$
approaches one of its two limiting values in
(\ref{eq:c-inequality-uniaxial-a}). For $cf= \frac{6 }{25 }d^2 $,
Fig.~\ref{fig:df-for-c-eq-1}, the $N_{U-}-N_{U-}$ transition line
and a part of the $N_{U+}-N_{U-}$ transition line become reduced to
a \emph{common straight line}
\begin{equation}\label{eq:Nu-Nu-Nu-limit}
    b= -\frac{5  f}{2 d}\,a, \hspace{1cm} a\in[-\frac{8
d^4}{625 f^3},0].
\end{equation}
That is, the $N_{U-}-N_{U-}$ transition line becomes also a line of
triple points with the critical and  triple  point collapsing at
$a=-\frac{8 d^4}{625 f^3}$! This case, illustrated in
Fig.~\ref{fig:nu-nu-nu-iso-degenerate}, makes us to expect that when
higher orders in the expansion (\ref{eq:degl0}) are taken into
account the degeneracy of the $N_{U-}-N_{U-}-N_{U+}$ line should be
removed and replaced by an $(Iso)-N_{U-}-N_{U-}-N_{U+}$
bubble-shaped diagram with up-to three triple points. Bracket
indicates that the branch $Iso-N_{U-}$ does not need to be present.
At $cf = \frac{9}{25 }d^2 $, which is the second  of the two
limits,
the $N_{U-}-N_{U-}$ line becomes reduced to a
single critical point located at
$(a,b)=(\frac{2 d^4}{625 f^3},-\frac{7 d^3}{125
   f^2}) $. The point belongs to the $Iso-N_{U-}$ transition line.

The phase diagrams described so far are stable against formation of
the biaxial phase given that $\phi\equiv f-1 \le 0$. If this
condition is fulfilled we can always find a uniaxial state with free
energy lower than- or equal to the free energy of any  biaxial
state. Indeed, consider a biaxial phase of $0< \omega <1$. A
sufficient condition for the equilibrium value $\omega_b$ of
$\omega$ in the biaxial phase is then given by
\begin{equation}\label{eq:biax-omega-sufficient}
\left. \frac{\partial F}{\partial \omega}\right|_{\omega_b} = F_b(q)
+ \frac{1}{3}\phi \,q^6{\omega_b}=0, \hspace{5mm} 0 < \omega_b < 1,
\end{equation}
which, when solved for $F_b(q)$ and substituted back to the biaxial
free energy formula, Eq.~(\ref{eq:degl0}), yields
\begin{equation}\label{eq:biax-free-energy}
  F_{biax}\equiv  F(q,\omega_b)= F_0+F_u(q)-\frac{1}{6}\phi\,
   q^6{\omega_b}^2.
\end{equation}
The Eq.~(\ref{eq:biax-free-energy}) clearly shows that only for
$\phi>0$ ($f>1$) there is a chance to get a stable biaxial nematic
phase. For $\phi < 0$  the uniaxial state is always more favorable.
The same conclusions are drawn for the biaxial state of $\omega=1$.
By a direct calculation of the  free energy we find for this case
that the biaxial state of the free energy $F(q=q_b,\omega=1)$, is
always less stable than one of the two uniaxial states: $\{q=\pm
q_b$, $\omega=0\}$, where $q_b$ is value of $q$ in the biaxial phase.
\subsection{Phase diagrams with biaxial nematic phase:
$q\ne 0, 0<\omega \le 1$}
The discussion of the previous section shows that, generally, a
stable biaxial nematic phase is found  for $\phi= f-e \equiv f-1
>0$. In this section we analyze this case more thoroughly. We start by
pointing out  that the sign of the $F_b(q)$ term in
Eq.~(\ref{eq:degl0}) decides about the relative stability of the
biaxial order with respect to all other phases involved. Generally,
a uniaxial phase becomes  unstable against formation of the
long-range biaxial order if  $F_b(q) \le 0$, which implies that
\begin{equation}\label{eq:instability-uni-vs-biax}
    b\, \gtrless \frac{3}{5}\, d \,q^2 + \phi\, q^3, \hspace{0.5cm}
    q\gtrless 0.
\end{equation}
In addition, for $F_b(q) \le -\frac{1}{3}\,\phi\,q^6$, or,
equivalently
\begin{equation}\label{eq:condition-wb-maximal}
    b\, \lessgtr \frac{3}{5}\, d \,q^2,  \hspace{0.5cm}
    q\gtrless 0
\end{equation}
the phase biaxiality, $\omega$, attains
its maximal value $\omega_b=1$, Eq.~(\ref{eq:biax-omega-sufficient}). The equality sign in the condition
(\ref{eq:instability-uni-vs-biax}) marks a  bifurcation from the
uniaxial to the biaxial phase. Together  with
(\ref{eq:uniaxial-necessary}), this can be  solved  for $a$ and $b$
to give the spinodal lines in a parametric form:
\begin{equation}\label{eq:bif-nu-biax}
    (a,b)=\left(-c q^2-\frac{2 d }{5}\,q^3 -q^4, \frac{3 d }{5}\,q^2 +
    \phi\,
   q^3\right).
\end{equation}

A few general conclusions can be drawn from the formulas
(\ref{eq:degl0},\ref{eq:bif-nu-biax}) and inequality
(\ref{eq:instability-uni-vs-biax}). First of all, if
Eq.~(\ref{eq:bif-nu-biax}) is fulfilled on a globally stable
uniaxial nematic branch,  the transition $N_U-N_B$ is second order.
Satisfying relation (\ref{eq:bif-nu-biax}) on a locally stable
uniaxial branch results in a first order $N_U\,(Iso)-N_B$ phase
transition. That is, the bifurcation scenario allows for a
possibility of a tricritical point on the $N_U-N_B$ line. Second
order $Iso-N_B$ transition is only admitted to states of maximal
biaxiality ($\omega=1$).

For the biaxial branch of the free energy a more quantitative
analysis can be given. In particular, the biaxial free energy
(\ref{eq:biax-free-energy}) can be expressed in an equivalent form
as
\begin{equation}\label{eq:free-energy-biax-q}
    F_{biax}=-\frac{b^2}{6 \phi }
+ \frac{1}{2} \alpha q^2 + \frac{1}{4}\gamma q^4 + \frac{q^6}{6},
\end{equation}
where
\begin{equation}\label{eq:biax-parameters}
    \alpha =a+\frac{2 b d}{5 \phi }, \hspace{5mm}\gamma
   =c -\frac{6\, d^2}{25 \phi } \hspace{4mm} \text{and} \hspace{4mm}
    \gamma ^2\geq 4 \alpha.
\end{equation}
A convenient parametric form for the $Iso-N_B$ line now easily
follows from the equation $F_{biax}=0$, supplemented with the
condition for $q_b$: $(\partial F_{biax}/\partial q)_{q=q_b}=0$.
 The solution of practical importance may be expressed as
\begin{equation}\label{eq:ab-biaxial-from-free-energy}
    \left( a = -\frac{2 b d}{5 \phi }-\gamma  q^2-q^4,
    b^2 = -\frac{1}{2} q^4 \left(4 q^2+3 \gamma \right)
    \phi\right).
\end{equation}
Additionally,  a stability criterion of the biaxial solution is
given by the condition that determinant of second derivatives of the
free energy is positive definite. This means that the biaxial phase
is  locally stable if
\begin{eqnarray}\label{eq:determinant-of second-derivs-biax}
    \frac{\partial ^2F}{\partial q^2} \frac{\partial
   ^2F}{\partial \omega ^2}-\left(\frac{\partial
   ^2F}{\partial q\, \partial \omega }\right)^2 \ge 0 \hspace{1cm}
   \Longrightarrow\nonumber
   \\
4 \alpha +\gamma  \left(\sqrt{\gamma ^2-4 \alpha }-\gamma \right)>
0.
\end{eqnarray}
The limiting case of vanishing determinant gives  two straight lines
in the $(a,b)$-plane: $\left\{ \alpha=0, \gamma^2=4\alpha \right\}$,
which are further  spinodals of the model.

Detailed  analysis of relative stability of $Iso$, $N_U$ and $N_B$
phases shows that all 'uniaxial' phase diagrams,
Figs.~\ref{fig:nu-iso}-\ref{fig:nu-nu-nu-iso-degenerate}, have their biaxial counterparts. Generally,  the biaxial phase
replaces, at least partly, the $N_{U+}-N_{U-}$ transition line by
the two lines: ${N_{U+}}\leftrightarrow{~N_B}$ and
${N_{U-}}\leftrightarrow{N_B}$. They can be given in a parametric
form as functions of the real parameter $q_1$
\begin{eqnarray}\label{eq:ab-Nu-Nb-parametric}
  {5 \delta a}{} &=& q_1\left[\left(6 q^2-10 q_1^2\right) d^2-10 \zeta
  q_1 d+ \right.\nonumber  \\&& \hspace{10mm}\left. 25 q^2
  \left(q^2+c\right) \phi
  \right] \\
5 \delta b &=&6 d^2 q^2-25 \phi  \left[q^4-q_1^3 \left(d+f
q_1\right)+ \right. \nonumber \\ &&\hspace{10mm}\left.
c \left(q^2-q_1^2\right)\right] \\
50 q^2 \phi &=&4 d^2-40 \phi  q_1 d-50 \zeta  \phi \pm
\nonumber\\
&&\hspace{-5mm}\sqrt{2} \sqrt{\delta ^2 \left(2 d^2+40 \phi  q_1
d+25
   (2 \zeta -c) \phi \right)},
\end{eqnarray}
with $\delta =2 d+5 \phi  q_1$  and  $\zeta =f q_1^2+c$.
The parameter $q_1$ runs over  the uniaxial branches, where $q_1>0$
for $N_{U+}$ and $q_1<0$ for $N_{U-}$. Analyzing various cases we
are able to single out six additional classes of the diagrams, shown
in Figs.~\ref{fig:nu-nb-landau-iso-d-eq-0}-\ref{fig:nu-nb-iso-i},
that supplement the uniaxial family. Again, the phase diagrams with
the $N_B$ phase should be correlated with
Figs.~\ref{fig:df-for-c-eq-m1}-\ref{fig:df-for-c-eq-1}, where
sectors of absolute stability of a given class  are shown in the
$\{c,d,f\}$ parameter space.

The discussion of the biaxial phase diagrams will proceed in a
similar way as for the uniaxial case, that is  we again start with
the case of $d=0$. In this limit we can distinguish between the two
different classes of the diagrams, all  being symmetric with respect
to the $b=0$ line.
\subsubsection{Class (e)}
\begin{figure}
\includegraphics*[width=8cm,height=8cm]{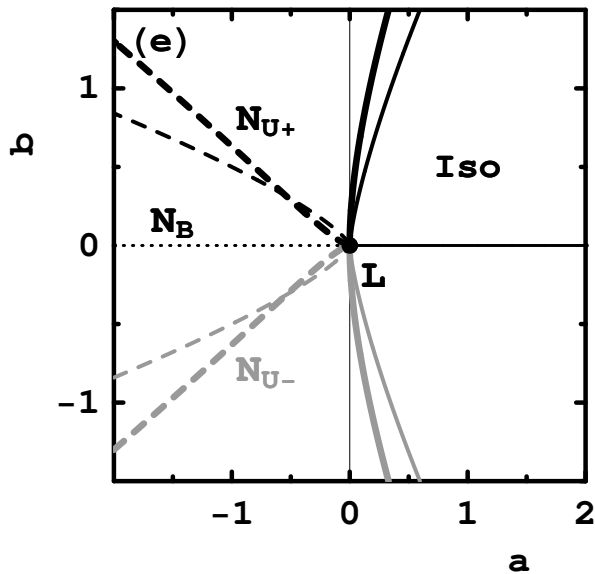}
\caption{Generic phase diagram for $d=0$ and $c \ge 0$. Biaxial
phase is sandwiched  between  two uniaxial phases. A direct phase
transition from the isotropic phase to biaxial nematic  is possible
through the Landau point. Solid lines represent phase transitions of
first order, dashed lines second order. Two cases are shown: $c=0, d
= 0, f = 1.5$ (thin lines)  and $c=1,d = 0, f = 1.5$ (thick lines).
As previously, gray lines represent phase transitions involving
$N_{U-}$ phase. \label{fig:nu-nb-landau-iso-d-eq-0}}
\end{figure}

The first class, shown in Fig.~\ref{fig:nu-nb-landau-iso-d-eq-0},
is similar to (a), Fig.~\ref{fig:nu-iso}.
The only difference is
that  the line
separating ${N_{U+}}$ and ${N_{U-}}$  splits itself into
${N_{U+}}\leftrightarrow{~N_B}$ and  ${N_{U-}}\leftrightarrow{N_B}$ lines
of second order phase transitions with $N_B$ phase positioned in between.
We find this class stable for $c \ge 0$ and
$f>1(\equiv\phi>0)$. As previously  the uniaxial lines are given by
 (\ref{eq:isoNuFord_0}). For the ${N_{U}}-{~N_B}$ lines
the formulas (\ref{eq:ab-Nu-Nb-parametric}) now simplify to
\begin{equation}\label{eq:Nu-Nb-lines-for-d-0}
    b^2=\frac{1}{2} \left(3 a c -c^3+\sqrt{c^2-4 a}
    \left(c^2-a\right)\right) \phi ^2,
\end{equation}
where $a\le0$.
 The four phases: $~Iso$, ${N_{U+}}$, ${N_B}$ and ${N_{U-}}$
meet at the Landau (tetracritical) point: $L = (a,b)=(0,0)$.
Additionally, for $c=0$  the ${N_{U+}}-{~N_B}$ and
${N_{U-}}-{~N_B}$ lines have  a common tangent at L, which is given
by the $a=0$-line. For $c=1$ this tangent is the $b=0$-line.
\subsubsection{Class (f)}
\begin{figure}
\includegraphics*[width=8.5cm,height=8.5cm]{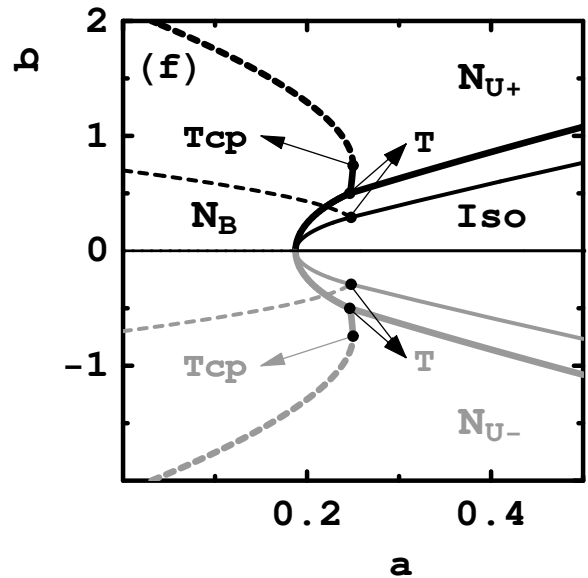}
\caption{Generic phase diagram for $d=0$ and $c < 0$. Biaxial phase
becomes stable  between  two uniaxial phases. A direct phase
transition from isotropic phase to biaxial nematic  is possible
along the line between the two triple points, T, that replace the Landau
point. Solid lines represent phase transitions of first order,
dashed lines second order. Tcp stands for tricritical point. Two
cases with $f\lessgtr2$ are shown: $c=-1, d = 0, f = 1.7$ (thin
lines) and $c=1,d = 0, f = 3.1$ (thick lines). For the meaning of
lines see caption to Fig.~\ref{fig:nu-nb-landau-iso-d-eq-0}.
 \label{fig:nu-nb-iso-split-d-eq-0}}
\end{figure}

The diagrams of this class,
Fig.~\ref{fig:nu-nb-iso-split-d-eq-0}, are  also derived from (a) and observed  when $c<0$.
Again  the uniaxial lines are given by
Eq.~(\ref{eq:isoNuFord_0}) and  the ${N_{U}}-{~N_B}$ lines by
Eq.~(\ref{eq:Nu-Nb-lines-for-d-0}). The latter represent
thermodynamically stable, second-order transition lines if
\begin{eqnarray}\label{eq:Nu-Nb-conditions-for-d-0}
  a &\le& \frac{3 c^2 (2 f-1)}{4 (f+1)^2} \hspace{5mm}\text{for}
  \hspace{5mm}1<f\le 2   \\
  a &\le& \frac{c^2}{4}\hspace{19mm} \text{for} \hspace{5mm} f > 2.
\end{eqnarray}
A new feature shown is a splitting of the Landau point into two
triple points where $Iso$, $N_{U}$ and $N_B$  meet. The position of
the triple points is
\begin{eqnarray}\label{eq:Nu-Nb-T-points-d-0}
  (a,b^2) &=& \left(\frac{3 c^2 (2 f-1)}{4 (f+1)^2},
    -\frac{27 c^3 (f-1)^2}{8 (f+1)^3}\right).
\end{eqnarray}
Both triple points are connected by a direct
$Iso\leftrightarrow{N_B}$ line of first order phase transitions  for
which we have
\begin{eqnarray}\label{eq:Iso-Nb-line-d-0}
    b^2=\frac{1}{4}\phi \left[
  c \left(c^2-6 a\right)-  \left(c^2-4 a\right)^{3/2}\right].
\end{eqnarray}
Depending on $f$, the phase transition between $N_U$ and $N_B$ can be
either first or second order.   For $1\le f \le 2$ only second order
$N_U-N_B$ transitions  are realized.  For $f>2$ the second order
transition line (\ref{eq:Nu-Nb-lines-for-d-0}) is separated from the
triple point by the   ${N_{U}}-{~N_B}$ line of first order phase
transitions ($\omega \ne 0$). Both  ${N_{U}}-{~N_B}$ lines meet at
the tricritical point given by
\begin{eqnarray}\label{eq:Nu-Nb-tricritical-point-for-d-0}
  (a,b^2) = \left(\frac{c^2}{4}, -\frac{1}{8} c^3 \phi ^2\right).
\end{eqnarray}
\subsubsection{Class (f')}

\begin{figure}
\includegraphics*[width=8.5cm,height=8.5cm]{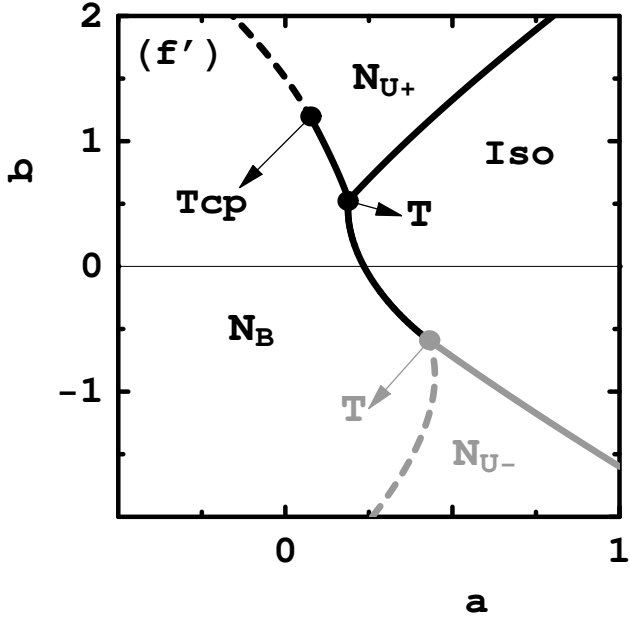}
\caption{Generic phase diagram for $d \ne 0$ and $c < 0$. Biaxial
phase becomes stable  between  two uniaxial phases. A direct phase
transition from the isotropic phase to biaxial nematic  is possible
along the line between two triple points (T) that replace the Landau
point. Only one tricritical point (Tcp), along $N_{U+}-N_B$,  is
possible for this class of diagrams.  Parameters taken are: $c=-1,d
= 1, f = 3$. For the meaning of lines see caption to
Fig.~\ref{fig:nu-nb-landau-iso-d-eq-0}. \label{fig:nu-nb-iso-fp}}
\end{figure}

\begin{figure}
\includegraphics*[width=8.5cm,height=8.5cm]{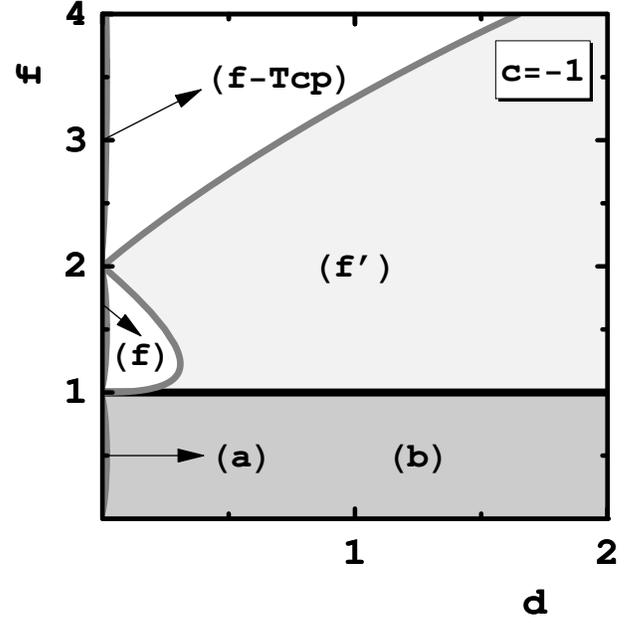}
\caption{ Sectors, in $\{c=-1,d,f\}$ parameter space, of absolute
stability  of phase diagrams labeled from (a) through (i). (f-Tcp)
stands for (f)-class with two tricritical points,
Fig.~\ref{fig:nu-nb-iso-split-d-eq-0}. Deformed versions of (f) and
(f-Tcp) diagrams are realized within sectors marked white.
\label{fig:df-for-c-eq-m1}}
\end{figure}

Now we turn to a more complex  case, namely that of $d\ne0$. It is
quite convenient to discuss new features of the diagrams that emerge
in  this case  by referring  directly to the parameter space
division as shown in
Figs.~\ref{fig:df-for-c-eq-m1}-\ref{fig:df-for-c-eq-1}. New classes
will be parameterized by  $c=\pm 1,0$.
 It
turns out that for $c=-1$,    the effect of nonzero $d$  is merely
to distort the phase diagrams classified as (f). The distorted
diagrams that preserve all features of (f) are separated from the
new class (f'), Fig.~\ref{fig:nu-nb-iso-fp}, by curves:
\begin{equation}\label{eq:f-from-fp}
  d^2 = \frac{25 c \left(\sqrt{\phi }-1\right)^2 \phi }{6
   \left(\phi -2 \sqrt{\phi }-2\right)}, \hspace{3mm} 0 \le \phi \le
   4+2\sqrt{3}.
\end{equation}
The curves are pictured dark-gray in Fig.~\ref{fig:df-for-c-eq-m1}.
The area to the right, shown as  light-gray, represents the class
(f'). The class differs from the deformed versions of (f-Tcp)-like
diagrams with two tricritical points and of (f) without tricritical
points by the presence of one tricritical point on the $N_{U+}-N_B$
transition line.
\subsubsection{Class (g)}
\begin{figure}
\includegraphics*[width=8.5cm,height=8.5cm]{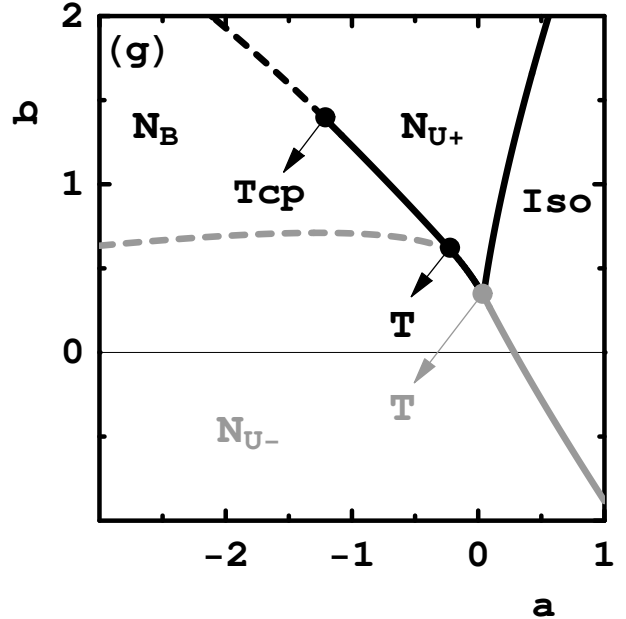}
\caption{Generic phase diagram for $d \ne 0$ and $c = 0$. Biaxial
phase becomes stable  between  two uniaxial phases. A direct phase
transition from isotropic phase to biaxial nematic  is not
possible. Two  triple points (T) are connected by the
$N_{U+}-N_{U-}$ line of first order phase transitions. One
tricritical point (Tcp) appears on the $N_{U+}-N_B$ line. Maximum
along $N_{U-}-N_B$ allows for two  biaxial nematic phases on the
temperature scale, separated by the $N_{U-}$ phase. The
low-temperature biaxial phase is referred to as reentrant $N_B$.
Parameters taken are: $c=0,d = 2, f = 1.6$. For the meaning of lines
see caption to Fig.~\ref{fig:nu-nb-landau-iso-d-eq-0}.
\label{fig:nu-nb-iso-g}}
\end{figure}

\begin{figure}
\includegraphics*[width=8.5cm,height=8.5cm]{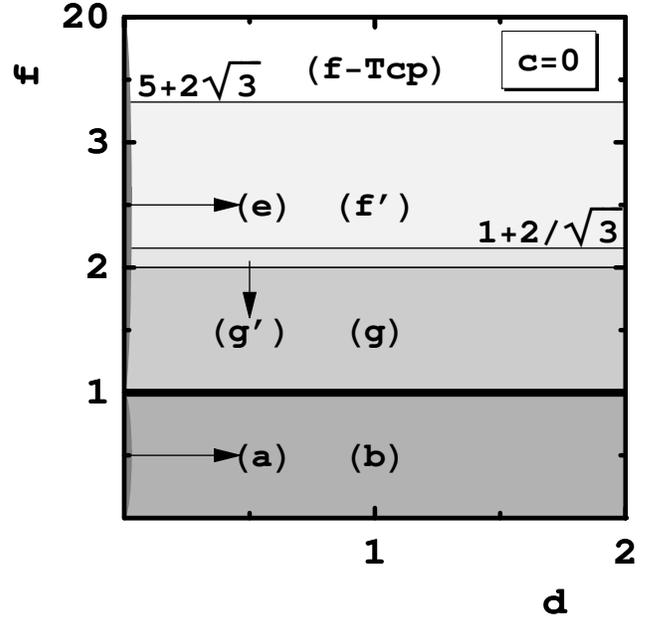}
\caption{Sectors, in $\{c=0,d,f\}$ parameter space, of absolute
stability  of phase diagrams labeled from (a) through (i). (g')
stands for  (g)-class without reentrant $N_B$ ({\em i.e.} without a
maximum along $N_{U-}-N_B$).
 \label{fig:df-for-c-eq-0}}
\end{figure}

The case $c=0$, Fig.~\ref{fig:df-for-c-eq-0}, results in a new class
of the diagrams shown in  Fig.~\ref{fig:nu-nb-iso-g}.
One of the
differences between (g) and (f'), exemplified in
Fig.~\ref{fig:nu-nb-iso-g}, is the absence of the direct transition
between $Iso$ and $N_B$. The
$N_B$ phase  branches off the $N_{U+}-N_{U-}$ first-order transition
line at the $N_{U+}-N_B-N_{U-}$ triple point. Interestingly, we observe a
maximum along the second-order  $N_B-N_{U-}$ transition line at the
location given by
\begin{equation}\label{eq:maximum-reentrant-g}
    (a,b)=\left( \frac{16d^4(\phi-1)}{625\phi^4},
    \frac{4d^3}{125\phi^2}
     \right).
\end{equation}
This maximum indicates that we can observe  reentrant biaxial
nematic phase as temperature is lowered. Consequently, it leads to a
very rich sequence of phase transitions, {\emph{e.g.}}
$Iso-N_{U+}-N_B-N_{U-}-N_B$. The reentrant phase and hence also the
maximum disappear for $f\ge 2$. In the interval  $2<f<
1+2/\sqrt{3}$, shown as sector (g') in Fig.~\ref{fig:df-for-c-eq-0},
the remaining features of the diagram, Fig.~\ref{fig:nu-nb-iso-g},
are left unchanged.
\subsubsection{Class (h)}
\begin{figure}
\includegraphics*[width=8.5cm,height=8.5cm]{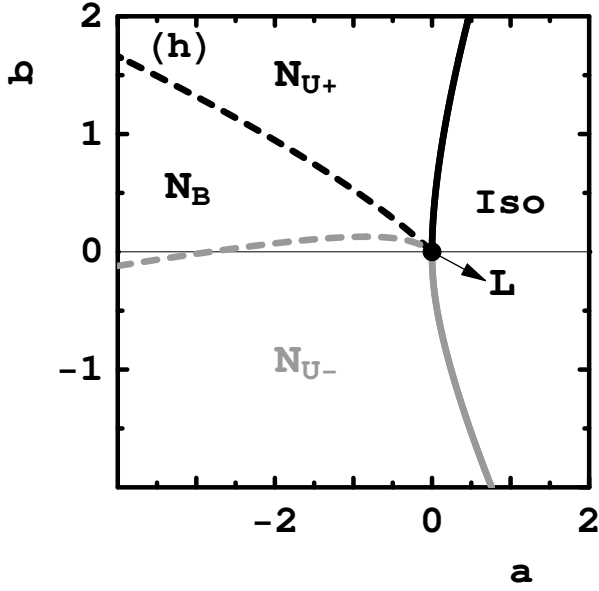}
\caption{Generic phase diagram for $d \ne 0$ and $c = 1$. This class
of the diagrams is a deformed versions of (e). A  major difference
between (e) and (h) is a maximum along $N_{U-}-N_B$ that allows for
reentrant $N_B$. Parameters taken are: $c=1,d = 1, f = 1.5$. For the
meaning of lines and of reentrant $N_B$ see captions to
Figs.~\ref{fig:nu-nb-landau-iso-d-eq-0} and \ref{fig:nu-nb-iso-g},
respectively. \label{fig:nu-nb-iso-h}}
\end{figure}

For $c=1$ we identify two new classes of the diagrams, denoted (h)
and  (i). The class (h),  Fig.~\ref{fig:nu-nb-iso-h}, is derived
from (e), the difference again being the presence of maximum along the second-order  $N_B-N_{U-}$
transition line at
$(a,b)$ given by Eq.~(\ref{eq:maximum-reentrant-g}).
That is we again can observe a sequence of phases
with reentrant biaxial nematic. Sector (h),
Fig.~\ref{fig:df-for-c-eq-1}, is separated from the neighboring
sectors (g) and (h+c) by the following lines: the dashed one given
by $f=1+6d^2/25$ ($0\le d \le 5/\sqrt{3}$) and the continuous one
given by $f=9d^2/25$ ($d > 5/\sqrt{3}$).
\subsubsection{Class (i)}
\begin{figure}
\includegraphics*[width=8.5cm,height=8.5cm]{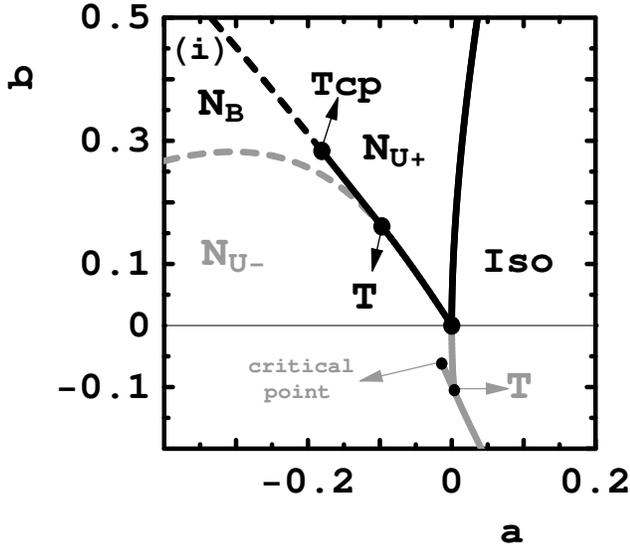}
\caption{Generic phase diagram for $d \ne 0$ and $c = 1$. It combines
properties  of (g) and (c). Parameters taken are: $c=1,d = 3, f =
2.75$. \label{fig:nu-nb-iso-i}}
\end{figure}

\begin{figure}
\includegraphics*[width=8.5cm,height=8.5cm]{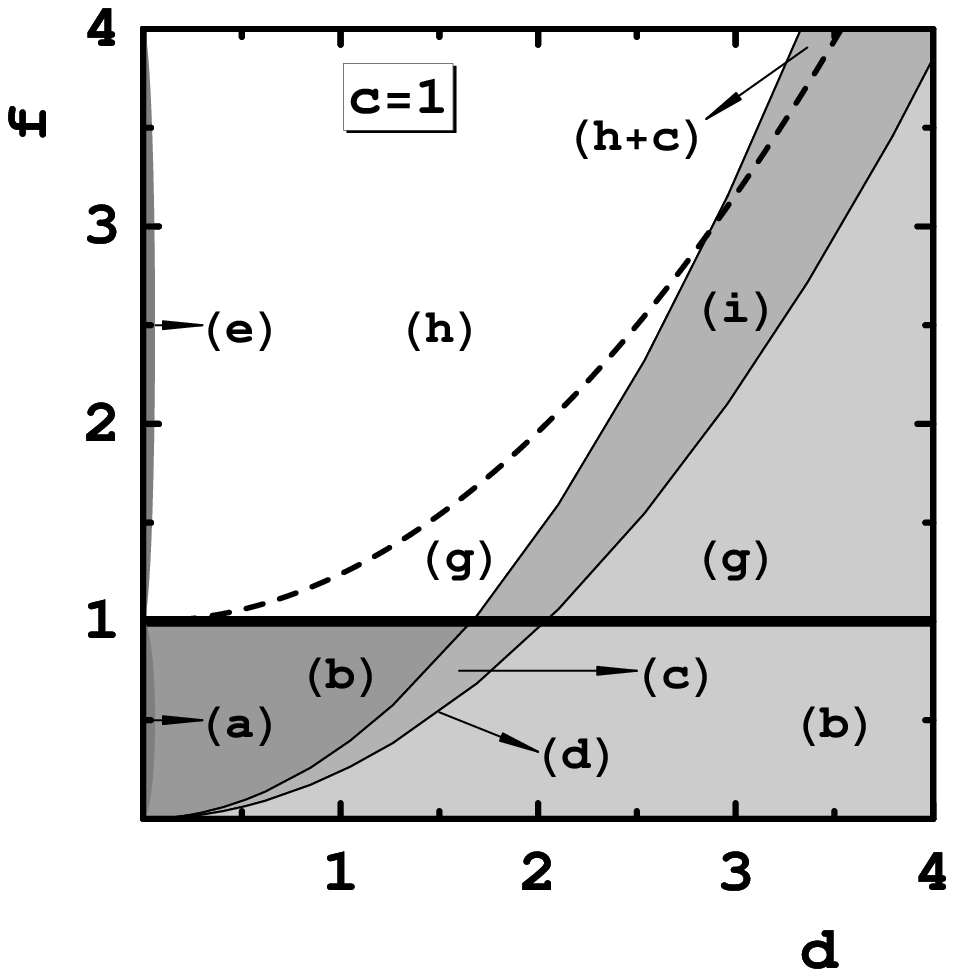}
\caption{Sectors, in $\{c=1,d,f\}$ parameter space, of absolute
stability  of phase diagrams labeled from (a) through (i). (h+c)
stands for  diagrams that combine properties of (h) and (c)
(\emph{see }Figs.~\ref{fig:nu-nu-nu-iso},\ref{fig:nu-nb-iso-h}).
\label{fig:df-for-c-eq-1}}
\end{figure}

This class of the diagrams is essentially a combination of (g) and
(c) and yields the richest sequences of phases and of corresponding
phase transitions. They include reentrant biaxial nematic,
$N_{U+}-N_B$ tricritical point and a line of phase transitions
between identical uniaxial phases terminating at a critical point.
Exemplary phase diagram is given in  Fig.~\ref{fig:nu-nb-iso-i}.
Sector of stability for this class, denoted (i) in
Fig.~\ref{fig:df-for-c-eq-1}, is limited by the following curves:
$f=1+6d^2/25$ ($d > 5/\sqrt{3}$), $f=9d^2/25$ ($5/3\le d \le
5/\sqrt{3}$), $f=1$ ($5/3\le d \le 5/\sqrt{6}$) and $f=6d^2/25$ ($d
> 5/\sqrt{6}$). In a small sector, named (h+c), the tricritical  point
and the $N_{U+}-N_{U-}$ line disappear, the resulting phase diagrams
being a combination of (h) and (c).
\subsection{Degenerated phase diagrams for b=0}
This case requires a few comments for not all phase sequences with b=0 can be identified from the diagrams that we have given so far. There are also subtle symmetry issues due to the presence of accidental degeneracy.

The identification of phases and of phase sequences for $b=0$
is straightforward for $d \ne 0$, in which case they follow directly from the diagrams representing classes (b,c,d,f',g,h). Only the case $d=0$, represented by the diagrams (a,e,f), requires separate comments.
By inspecting the free energy expansion (\ref{eq:degl0}) for $b=d=0$ we find immediately that only three states can be realized at equilibrium: (\textit{i}) degenerated uniaxial state of $\omega=0$ for $f<1$; (\textit{ii}) degenerated biaxial-uniaxial state of arbitrary $\omega$ for $f=1$, and (\textit{iii}) biaxial state of maximal biaxiality ($\omega=1$) for $f>0$.

The uniaxial state, denoted (\textit{i}), has the same free energy for oblate and prolate states, which means that the system creates oblate and prolate domains with the same energy cost. Due to this accidental degeneracy the symmetry group of the state can be classified as ${\cal{D}}_{\infty h}\times {\cal{Z}}_{2}$. Consequently,  the transition from the isotropic phase to the degenerated uniaxial phase can be either second-order ($c=1$) or first order ($c=-1$) with a tricritical point at $c=0$. The transition temperature and the order
parameter for temperatures below the transition are given by $\{ a=0$ \emph{for} $c=0,1$;
 $a=\frac{3 }{16 f}$ \emph{for} $c=-1\} $ and $q^2=\frac{\sqrt{c^2-4 a f}-c}{2 f}$, respectively.

The states (\textit{ii},\textit{iii}) have mathematically the same form of the free energy as one for the degenerated uniaxial case. All formulas are reproduced from the case
(\textit{i}) if we substitute $f=1$ there.
Hence, again, the phase transition from the isotropic phase to the corresponding ordered phase can be either first or second  order with an intermediate tricritical point.
The difference between the cases (\textit{ii}) and (\textit{iii}) is in symmetry. For (\textit{ii})
the only restriction on $\mathbf{Q}$  is $\mathrm{Tr}{(\mathbf{Q}}^{2})=const$. The accidental degeneracy of the equilibrium solutions for $\mathbf{Q}$ is the rotational invariance
in five-dimensional space of components $Q_{\alpha\beta}$. The relevant symmetry group is  thus ${\cal{O}}({5})$, which seizes up the difference between uniaxial and biaxial domains. We call it the degenerated biaxial-uniaxial phase.

In the case (\textit{iii}) the tensor $\mathbf{Q}$ is given by Eq.~(\ref{eq:Q-diagonal}) with {\emph{e.g.}} $q_0=0$, that is by one of the three degenerated solutions fulfilling maximal biaxiality condition $\mathrm{Tr}{(\mathbf{Q}}^{3})=0$.
\section{Discussion}
The Landau-deGennes theory of biaxial nematics presented in this paper has been elaborated to show the outcome of  mathematical structure of the  expansion that is based solely on the order parameter $\mathbf{Q}$. The  phenomenological approach is particularly simple and the knowledge of full spectrum of predictions of one of the most commonly
cited theories is desirable, particularly because of current interest in seeking for stable thermotropic biaxial nematic phase. The analytical formulas are given for almost all transition lines, characteristic points of the lines, and for the stability range of a given class of the phase diagrams.  Except for
the purely uniaxial group of the diagrams for $f<1$, the biaxial
phase is naturally stabilized between prolate and oblate uniaxial
nematics. Phase transitions to the biaxial phase can be either first
or second order with a possibility of a tricritical point. Due to
a competition between cubic and fifth-order invariants the direct $N_{U-}-N_{U-}$, $N_{U+}-N_{U-}$ and the reentrant
biaxial nematic phase are also possible. The Landau $Iso-N_{U+}-N_B-N_{U-}$ (tetracritical) point can split into two triple points positioned either on the $Iso-N_{B}$ transition line or on the $N_{U+}-N_{U-}$ line.

One may wonder then why the
biaxial nematic phase is so difficult to account for experimentally.
The practical difficulty could be  that for real systems the
 parameter $f$, responsible for the stabilization of the  biaxial nematic phase, is much too small compared to other
coefficients of the Landau expansion, so we effectively stay  in the
uniaxial sector of the diagrams ($f\le 1$). An alternative explanation could be
that smectic and crystalline phases, not taken into account, may interfere before the right
thermodynamic parameters are reached.  Clearly, the best choice of
the Landau coefficients to get $N_B$ absolutely stable would be that  where  $N_B$ bifurcates directly from the isotropic
phase. At microscopic level it would then be of interest to construct molecular models showing generic types of the diagrams identified phenomenologically. To this goal it is necessary to establish a bridge between molecular and
phenomenological approaches, in particular one needs a molecular interpretation of the alignment tensor and, at least,  of the $(a,b)$-parameters entering the expansion. The problem is relatively simple in the mean-field approximation and the solution has already been given \cite{Mulder:pra:biaxBif:1989,Longa:lc:Qtensor:MolInterpretation2005} for the class of the so called $L=2$ models with ${\cal{D}}_{2 h}$-symmetric hard molecules/soft interactions \cite{Mulder:pra:biaxBif:1989,Longa:pre:biax:2005}.
Applying formulas (14-20) from \cite{Longa:lc:Qtensor:MolInterpretation2005} to the mean-field versions of the models \cite{Alben:prl:biax:1973,Luckhurst:mp:biax:dispersionModel:1980,%
Sonnet:pre:biax:2003,Longa:pre:biax:2005,%
Longa:pre:biaxSU3:2007,Bates:pre:biax:FlexibleVshaped:2006}
we recover diagrams represented by (e) \cite{Alben:prl:biax:1973,Luckhurst:mp:biax:dispersionModel:1980,%
Longa:pre:biaxSU3:2007},
(f) \cite{Sonnet:pre:biax:2003,Longa:pre:biax:2005}
 and (g) \cite{Bates:pre:biax:FlexibleVshaped:2006}.
Evidence for degenerated states ($b=d=0$) has been given by Matteis and Virga \cite{Matteis:pre:biax:tcp:2005}.
\begin{acknowledgments}
This work was supported by Grant N202 169 31/3455 of Polish Ministry
of Science and Higher Education, and by the EC Marie Curie Actions 'Transfer of Knowledge', project COCOS (contract MTKD-CT-2004-517186).
\end{acknowledgments}
%

%

\bibliographystyle{apsrev}
\bibliography{BIAX-BIBLIOGRAPHY}

\begin{thebibliography}{38}
\expandafter\ifx\csname natexlab\endcsname\relax\def\natexlab#1{#1}\fi
\expandafter\ifx\csname bibnamefont\endcsname\relax
  \def\bibnamefont#1{#1}\fi
\expandafter\ifx\csname bibfnamefont\endcsname\relax
  \def\bibfnamefont#1{#1}\fi
\expandafter\ifx\csname citenamefont\endcsname\relax
  \def\citenamefont#1{#1}\fi
\expandafter\ifx\csname url\endcsname\relax
  \def\url#1{\texttt{#1}}\fi
\expandafter\ifx\csname urlprefix\endcsname\relax\def\urlprefix{URL }\fi
\providecommand{\bibinfo}[2]{#2}
\providecommand{\eprint}[2][]{\url{#2}}

\bibitem[{\citenamefont{Freiser}(1970)}]{Freiser:prl:biax:1970}
\bibinfo{author}{\bibfnamefont{M.~J.} \bibnamefont{Freiser}},
  \bibinfo{journal}{Phys. Rev. Lett.} \textbf{\bibinfo{volume}{24}},
  \bibinfo{pages}{1041} (\bibinfo{year}{1970}).

\bibitem[{\citenamefont{Freiser}(1971)}]{Freiser:mclc:biax:1971}
\bibinfo{author}{\bibfnamefont{M.~J.} \bibnamefont{Freiser}},
  \bibinfo{journal}{Mol. Cryst. Liq. Cryst.} \textbf{\bibinfo{volume}{14}},
  \bibinfo{pages}{165} (\bibinfo{year}{1971}).

\bibitem[{\citenamefont{Yu and Saupe}(1980)}]{Yu:prl:biax:1980}
\bibinfo{author}{\bibfnamefont{L.~J.} \bibnamefont{Yu}} \bibnamefont{and}
  \bibinfo{author}{\bibfnamefont{A.}~\bibnamefont{Saupe}},
  \bibinfo{journal}{Phys. Rev. Lett.} \textbf{\bibinfo{volume}{45}},
  \bibinfo{pages}{1000} (\bibinfo{year}{1980}).

\bibitem[{\citenamefont{Luckhurst}(2001)}]{Luckhurst:thinSolFilms:biax:rev:200%
1}
\bibinfo{author}{\bibfnamefont{G.~R.} \bibnamefont{Luckhurst}},
  \bibinfo{journal}{Thin Solid Films} \textbf{\bibinfo{volume}{393}},
  \bibinfo{pages}{40} (\bibinfo{year}{2001}).

\bibitem[{\citenamefont{Luckhurst}(2004)}]{Luckhurst:nature:biax:rev:2004}
\bibinfo{author}{\bibfnamefont{G.~R.} \bibnamefont{Luckhurst}},
  \bibinfo{journal}{Nature (London)} \textbf{\bibinfo{volume}{430}},
  \bibinfo{pages}{413} (\bibinfo{year}{2004}).

\bibitem[{\citenamefont{Madsen et~al.}(2004)\citenamefont{Madsen, Dingemans,
  Nakata, and Samulski}}]{Madsen:prl:biax:2004}
\bibinfo{author}{\bibfnamefont{L.~A.} \bibnamefont{Madsen}},
  \bibinfo{author}{\bibfnamefont{T.~J.} \bibnamefont{Dingemans}},
  \bibinfo{author}{\bibfnamefont{M.}~\bibnamefont{Nakata}}, \bibnamefont{and}
  \bibinfo{author}{\bibfnamefont{E.~T.} \bibnamefont{Samulski}},
  \bibinfo{journal}{Phys. Rev. Lett.} \textbf{\bibinfo{volume}{92}},
  \bibinfo{pages}{145505} (\bibinfo{year}{2004}).

\bibitem[{\citenamefont{Acharya et~al.}(2004)\citenamefont{Acharya, Primak, and
  Kumar}}]{Acharya:prl:biax:2004}
\bibinfo{author}{\bibfnamefont{B.~R.} \bibnamefont{Acharya}},
  \bibinfo{author}{\bibfnamefont{A.}~\bibnamefont{Primak}}, \bibnamefont{and}
  \bibinfo{author}{\bibfnamefont{S.}~\bibnamefont{Kumar}},
  \bibinfo{journal}{Phys. Rev. Lett.} \textbf{\bibinfo{volume}{92}},
  \bibinfo{pages}{145506} (\bibinfo{year}{2004}).

\bibitem[{\citenamefont{Merkel et~al.}(2004)\citenamefont{Merkel, Kocot, Vij,
  Korlacki, Mehl, , and Meyer}}]{Merkel:prl:biax:2004}
\bibinfo{author}{\bibfnamefont{K.}~\bibnamefont{Merkel}},
  \bibinfo{author}{\bibfnamefont{A.}~\bibnamefont{Kocot}},
  \bibinfo{author}{\bibfnamefont{J.~K.} \bibnamefont{Vij}},
  \bibinfo{author}{\bibfnamefont{R.}~\bibnamefont{Korlacki}},
  \bibinfo{author}{\bibfnamefont{G.~H.} \bibnamefont{Mehl}}, ,
  \bibnamefont{and} \bibinfo{author}{\bibfnamefont{T.}~\bibnamefont{Meyer}},
  \bibinfo{journal}{Phys. Rev. Lett.} \textbf{\bibinfo{volume}{93}},
  \bibinfo{pages}{237801} (\bibinfo{year}{2004}).

\bibitem[{\citenamefont{Madsen et~al.}(2006)\citenamefont{Madsen, Dingemans,
  Nakata, and Samulski}}]{Madsen:prl:biax:replyToGalerne2006}
\bibinfo{author}{\bibfnamefont{L.~A.} \bibnamefont{Madsen}},
  \bibinfo{author}{\bibfnamefont{T.~J.} \bibnamefont{Dingemans}},
  \bibinfo{author}{\bibfnamefont{M.}~\bibnamefont{Nakata}}, \bibnamefont{and}
  \bibinfo{author}{\bibfnamefont{E.~T.} \bibnamefont{Samulski}},
  \bibinfo{journal}{Phys. Rev. Lett.} \textbf{\bibinfo{volume}{96}},
  \bibinfo{pages}{219804} (\bibinfo{year}{2006}).

\bibitem[{\citenamefont{Neupane et~al.}(2006)\citenamefont{Neupane, Kang,
  Sharma, Carney, Meyer, Mehl, Allender, Kumar, and
  Sprunt}}]{Neupane:prl:lightScatteringTetrapodes:2006}
\bibinfo{author}{\bibfnamefont{K.}~\bibnamefont{Neupane}},
  \bibinfo{author}{\bibfnamefont{S.}~\bibnamefont{Kang}},
  \bibinfo{author}{\bibfnamefont{S.}~\bibnamefont{Sharma}},
  \bibinfo{author}{\bibfnamefont{D.}~\bibnamefont{Carney}},
  \bibinfo{author}{\bibfnamefont{T.}~\bibnamefont{Meyer}},
  \bibinfo{author}{\bibfnamefont{G.~H.} \bibnamefont{Mehl}},
  \bibinfo{author}{\bibfnamefont{D.}~\bibnamefont{Allender}},
  \bibinfo{author}{\bibfnamefont{S.}~\bibnamefont{Kumar}}, \bibnamefont{and}
  \bibinfo{author}{\bibfnamefont{S.}~\bibnamefont{Sprunt}},
  \bibinfo{journal}{Phys. Rev. Lett.} \textbf{\bibinfo{volume}{97}},
  \bibinfo{pages}{207802} (\bibinfo{year}{2006}).

\bibitem[{\citenamefont{de~Gennes and Prost}(1993)}]{DeGennes:book}
\bibinfo{author}{\bibfnamefont{P.~G.} \bibnamefont{de~Gennes}}
  \bibnamefont{and} \bibinfo{author}{\bibfnamefont{J.}~\bibnamefont{Prost}},
  \emph{\bibinfo{title}{The Physics of Liquid Crystals}}
  (\bibinfo{publisher}{Clarendon Press}, \bibinfo{address}{Oxford},
  \bibinfo{year}{1993}), \bibinfo{edition}{{Second}} ed.

\bibitem[{\citenamefont{Gramsbergen et~al.}(1986)\citenamefont{Gramsbergen,
  Longa, and de~Jeu}}]{Gramsbergen:pr:nem:1986}
\bibinfo{author}{\bibfnamefont{E.~F.} \bibnamefont{Gramsbergen}},
  \bibinfo{author}{\bibfnamefont{L.}~\bibnamefont{Longa}}, \bibnamefont{and}
  \bibinfo{author}{\bibfnamefont{W.~H.} \bibnamefont{de~Jeu}},
  \bibinfo{journal}{Phys. Rep.} \textbf{\bibinfo{volume}{135}},
  \bibinfo{pages}{195} (\bibinfo{year}{1986}).

\bibitem[{\citenamefont{Singh}(2000)}]{Singh:review:2000}
\bibinfo{author}{\bibfnamefont{S.}~\bibnamefont{Singh}},
  \bibinfo{journal}{Phys. Rep.} \textbf{\bibinfo{volume}{324}},
  \bibinfo{pages}{107} (\bibinfo{year}{2000}).

\bibitem[{\citenamefont{Longa et~al.}(1994)\citenamefont{Longa, Fink, and
  Trebin}}]{Longa:pre:biax:1994}
\bibinfo{author}{\bibfnamefont{L.}~\bibnamefont{Longa}},
  \bibinfo{author}{\bibfnamefont{W.}~\bibnamefont{Fink}}, \bibnamefont{and}
  \bibinfo{author}{\bibfnamefont{H.~R.} \bibnamefont{Trebin}},
  \bibinfo{journal}{Phys. Rev. E} \textbf{\bibinfo{volume}{50}},
  \bibinfo{pages}{3841} (\bibinfo{year}{1994}).

\bibitem[{\citenamefont{Charvolin}(1984)}]{Charvolin:nc:lyotr:biax:1984}
\bibinfo{author}{\bibfnamefont{J.}~\bibnamefont{Charvolin}},
  \bibinfo{journal}{Nuovo Cimento D} \textbf{\bibinfo{volume}{3}},
  \bibinfo{pages}{3} (\bibinfo{year}{1984}).

\bibitem[{\citenamefont{Neto and Salinas}(2005)}]{Figueiredo:book:lyotr:2005}
\bibinfo{author}{\bibfnamefont{A.~M.~F.} \bibnamefont{Neto}} \bibnamefont{and}
  \bibinfo{author}{\bibfnamefont{S.~R.~A.} \bibnamefont{Salinas}},
  \emph{\bibinfo{title}{The Physics of Lyotropic Liquid Crystals: Phase
  Transitions and Structural Properties}}, Monographs on the physics and
  chemistry of materials (\bibinfo{publisher}{Oxford University Press},
  \bibinfo{year}{2005}), ISBN \bibinfo{isbn}{0 19 85 2550}.

\bibitem[{\citenamefont{Alben}(1973)}]{Alben:prl:biax:1973}
\bibinfo{author}{\bibfnamefont{R.}~\bibnamefont{Alben}},
  \bibinfo{journal}{Phys. Rev. Lett.} \textbf{\bibinfo{volume}{30}},
  \bibinfo{pages}{778} (\bibinfo{year}{1973}).

\bibitem[{\citenamefont{Mulder}(1989)}]{Mulder:pra:biaxBif:1989}
\bibinfo{author}{\bibfnamefont{B.}~\bibnamefont{Mulder}},
  \bibinfo{journal}{Phys. Rev. A} \textbf{\bibinfo{volume}{39}},
  \bibinfo{pages}{360} (\bibinfo{year}{1989}).

\bibitem[{\citenamefont{Teixeira et~al.}(1998)\citenamefont{Teixeira, Masters,
  and Mulder}}]{Teixeira:mclc:biax:1998}
\bibinfo{author}{\bibfnamefont{P.~I.~C.} \bibnamefont{Teixeira}},
  \bibinfo{author}{\bibfnamefont{A.~J.} \bibnamefont{Masters}},
  \bibnamefont{and} \bibinfo{author}{\bibfnamefont{B.~M.}
  \bibnamefont{Mulder}}, \bibinfo{journal}{Mol. Cryst. Liq. Cryst.}
  \textbf{\bibinfo{volume}{323}}, \bibinfo{pages}{167} (\bibinfo{year}{1998}).

\bibitem[{\citenamefont{Sonnet et~al.}(2003)\citenamefont{Sonnet, Virga, and
  Durand}}]{Sonnet:pre:biax:2003}
\bibinfo{author}{\bibfnamefont{A.~M.} \bibnamefont{Sonnet}},
  \bibinfo{author}{\bibfnamefont{E.~G.} \bibnamefont{Virga}}, \bibnamefont{and}
  \bibinfo{author}{\bibfnamefont{G.~E.} \bibnamefont{Durand}},
  \bibinfo{journal}{Phys. Rev. E} \textbf{\bibinfo{volume}{67}},
  \bibinfo{pages}{061701} (\bibinfo{year}{2003}).

\bibitem[{\citenamefont{Longa et~al.}(2005)\citenamefont{Longa, Grzybowski,
  Romano, and Virga}}]{Longa:pre:biax:2005}
\bibinfo{author}{\bibfnamefont{L.}~\bibnamefont{Longa}},
  \bibinfo{author}{\bibfnamefont{P.}~\bibnamefont{Grzybowski}},
  \bibinfo{author}{\bibfnamefont{S.}~\bibnamefont{Romano}}, \bibnamefont{and}
  \bibinfo{author}{\bibfnamefont{E.~G.} \bibnamefont{Virga}},
  \bibinfo{journal}{Phys. Rev. E} \textbf{\bibinfo{volume}{71}},
  \bibinfo{pages}{051714} (\bibinfo{year}{2005}).

\bibitem[{\citenamefont{Longa et~al.}(2007)\citenamefont{Longa, Paj\c{a}k, and
  Wydro}}]{Longa:pre:biaxSU3:2007}
\bibinfo{author}{\bibfnamefont{L.}~\bibnamefont{Longa}},
  \bibinfo{author}{\bibfnamefont{G.}~\bibnamefont{Paj\c{a}k}},
  \bibnamefont{and} \bibinfo{author}{\bibfnamefont{T.}~\bibnamefont{Wydro}},
  \bibinfo{journal}{Phys. Rev. E} \textbf{\bibinfo{volume}{76}},
  \bibinfo{pages}{011703} (\bibinfo{year}{2007}).

\bibitem[{\citenamefont{Camp and Allen}(1997)}]{Camp:jcp:biax:1997}
\bibinfo{author}{\bibfnamefont{P.~J.} \bibnamefont{Camp}} \bibnamefont{and}
  \bibinfo{author}{\bibfnamefont{M.~P.} \bibnamefont{Allen}},
  \bibinfo{journal}{J. Chem. Phys.} \textbf{\bibinfo{volume}{106}},
  \bibinfo{pages}{6681} (\bibinfo{year}{1997}).

\bibitem[{\citenamefont{Camp et~al.}(1999)\citenamefont{Camp, Allen, and
  Masters}}]{Camp:jcp:biax:1999}
\bibinfo{author}{\bibfnamefont{P.~J.} \bibnamefont{Camp}},
  \bibinfo{author}{\bibfnamefont{M.~P.} \bibnamefont{Allen}}, \bibnamefont{and}
  \bibinfo{author}{\bibfnamefont{A.~J.} \bibnamefont{Masters}},
  \bibinfo{journal}{J. Chem. Phys.} \textbf{\bibinfo{volume}{111}},
  \bibinfo{pages}{9871} (\bibinfo{year}{1999}).

\bibitem[{\citenamefont{Cleaver et~al.}(1996)\citenamefont{Cleaver, Care,
  Allen, and Neal}}]{Cleaver:pre:biax:1996}
\bibinfo{author}{\bibfnamefont{D.~J.} \bibnamefont{Cleaver}},
  \bibinfo{author}{\bibfnamefont{C.~M.} \bibnamefont{Care}},
  \bibinfo{author}{\bibfnamefont{M.~P.} \bibnamefont{Allen}}, \bibnamefont{and}
  \bibinfo{author}{\bibfnamefont{M.~P.} \bibnamefont{Neal}},
  \bibinfo{journal}{Phys. Rev. E} \textbf{\bibinfo{volume}{54}},
  \bibinfo{pages}{559} (\bibinfo{year}{1996}).

\bibitem[{\citenamefont{Sarman}(1996)}]{Sarman:jcp:biax:1996}
\bibinfo{author}{\bibfnamefont{S.}~\bibnamefont{Sarman}}, \bibinfo{journal}{J.
  Chem. Phys.} \textbf{\bibinfo{volume}{104}}, \bibinfo{pages}{342}
  (\bibinfo{year}{1996}).

\bibitem[{\citenamefont{Ginzburg et~al.}(1997)\citenamefont{Ginzburg, Glaser, ,
  and Clark}}]{Ginzburg:cp:simBiax:1997}
\bibinfo{author}{\bibfnamefont{V.~V.} \bibnamefont{Ginzburg}},
  \bibinfo{author}{\bibfnamefont{M.~A.} \bibnamefont{Glaser}}, ,
  \bibnamefont{and} \bibinfo{author}{\bibfnamefont{N.~A.} \bibnamefont{Clark}},
  \bibinfo{journal}{Chem. Phys.} \textbf{\bibinfo{volume}{214}},
  \bibinfo{pages}{253} (\bibinfo{year}{1997}).

\bibitem[{\citenamefont{Berardi and Zannoni}(2000)}]{Berardi:jcp:biax:2000}
\bibinfo{author}{\bibfnamefont{R.}~\bibnamefont{Berardi}} \bibnamefont{and}
  \bibinfo{author}{\bibfnamefont{C.}~\bibnamefont{Zannoni}},
  \bibinfo{journal}{J. Chem. Phys.} \textbf{\bibinfo{volume}{113}},
  \bibinfo{pages}{5971} (\bibinfo{year}{2000}).

\bibitem[{\citenamefont{Biscarini et~al.}(1995)\citenamefont{Biscarini,
  Chiccoli, Pasini, Semeria, and Zannoni}}]{Biscarini:prl:biax:1995}
\bibinfo{author}{\bibfnamefont{F.}~\bibnamefont{Biscarini}},
  \bibinfo{author}{\bibfnamefont{C.}~\bibnamefont{Chiccoli}},
  \bibinfo{author}{\bibfnamefont{P.}~\bibnamefont{Pasini}},
  \bibinfo{author}{\bibfnamefont{F.}~\bibnamefont{Semeria}}, \bibnamefont{and}
  \bibinfo{author}{\bibfnamefont{C.}~\bibnamefont{Zannoni}},
  \bibinfo{journal}{Phys. Rev. Lett} \textbf{\bibinfo{volume}{75}},
  \bibinfo{pages}{1803} (\bibinfo{year}{1995}).

\bibitem[{\citenamefont{Allender and Lee}(1984)}]{Allender:mclc:biax:1984}
\bibinfo{author}{\bibfnamefont{D.~W.} \bibnamefont{Allender}} \bibnamefont{and}
  \bibinfo{author}{\bibfnamefont{M.~A.} \bibnamefont{Lee}},
  \bibinfo{journal}{Mol. Cryst. Liq. Cryst.} \textbf{\bibinfo{volume}{110}},
  \bibinfo{pages}{331} (\bibinfo{year}{1984}).

\bibitem[{\citenamefont{Allender et~al.}(1985)\citenamefont{Allender, Lee, and
  Hafiz}}]{Allender:mclc:biax:1985}
\bibinfo{author}{\bibfnamefont{D.~W.} \bibnamefont{Allender}},
  \bibinfo{author}{\bibfnamefont{M.~A.} \bibnamefont{Lee}}, \bibnamefont{and}
  \bibinfo{author}{\bibfnamefont{N.}~\bibnamefont{Hafiz}},
  \bibinfo{journal}{Mol. Cryst. Liq. Cryst.} \textbf{\bibinfo{volume}{124}},
  \bibinfo{pages}{45} (\bibinfo{year}{1985}).

\bibitem[{\citenamefont{Prostakov et~al.}(2002)\citenamefont{Prostakov, Larin,
  and Stryukov}}]{Prostakov:cr:biaxpheno:2002}
\bibinfo{author}{\bibfnamefont{A.~E.} \bibnamefont{Prostakov}},
  \bibinfo{author}{\bibfnamefont{E.~S.} \bibnamefont{Larin}}, \bibnamefont{and}
  \bibinfo{author}{\bibfnamefont{M.~B.} \bibnamefont{Stryukov}},
  \bibinfo{journal}{Cryst. Reports} \textbf{\bibinfo{volume}{47}},
  \bibinfo{pages}{1041} (\bibinfo{year}{2002}).

\bibitem[{\citenamefont{Longa et~al.}(1998)\citenamefont{Longa, Trebin, and
  {\.Z}elazna}}]{Longa:revBras:1998}
\bibinfo{author}{\bibfnamefont{L.}~\bibnamefont{Longa}},
  \bibinfo{author}{\bibfnamefont{H.-R.} \bibnamefont{Trebin}},
  \bibnamefont{and}
  \bibinfo{author}{\bibfnamefont{M.}~\bibnamefont{{\.Z}elazna}},
  \emph{\bibinfo{title}{{\em Phenomenological Approach to Phase Transitions in
  Complex Fluids} in {\em Phase Transitions in Complex Fluids}}}
  (\bibinfo{publisher}{(World Sci, Singapore)}, \bibinfo{address}{Singapore},
  \bibinfo{year}{1998}).

\bibitem[{\citenamefont{Toledano et~al.}(1995)\citenamefont{Toledano, Neto,
  Lorman, Mettout, and Dmitriev}}]{Toledano:pre:biaxdiag:1995}
\bibinfo{author}{\bibfnamefont{P.}~\bibnamefont{Toledano}},
  \bibinfo{author}{\bibfnamefont{A.~M.~F.} \bibnamefont{Neto}},
  \bibinfo{author}{\bibfnamefont{V.}~\bibnamefont{Lorman}},
  \bibinfo{author}{\bibfnamefont{B.}~\bibnamefont{Mettout}}, \bibnamefont{and}
  \bibinfo{author}{\bibfnamefont{V.}~\bibnamefont{Dmitriev}},
  \bibinfo{journal}{Phys. Rev. E} \textbf{\bibinfo{volume}{52}},
  \bibinfo{pages}{5040} (\bibinfo{year}{1995}).

\bibitem[{\citenamefont{Longa and
  Paj\c{a}k}(2005)}]{Longa:lc:Qtensor:MolInterpretation2005}
\bibinfo{author}{\bibfnamefont{L.}~\bibnamefont{Longa}} \bibnamefont{and}
  \bibinfo{author}{\bibfnamefont{G.}~\bibnamefont{Paj\c{a}k}},
  \bibinfo{journal}{Liq. Cryst.} \textbf{\bibinfo{volume}{32}},
  \bibinfo{pages}{1409} (\bibinfo{year}{2005}).

\bibitem[{\citenamefont{Luckhurst and
  Romano}(1980)}]{Luckhurst:mp:biax:dispersionModel:1980}
\bibinfo{author}{\bibfnamefont{G.~R.} \bibnamefont{Luckhurst}}
  \bibnamefont{and} \bibinfo{author}{\bibfnamefont{S.}~\bibnamefont{Romano}},
  \bibinfo{journal}{Mol. Phys.} \textbf{\bibinfo{volume}{40}},
  \bibinfo{pages}{129} (\bibinfo{year}{1980}).

\bibitem[{\citenamefont{Bates}(2006)}]{Bates:pre:biax:FlexibleVshaped:2006}
\bibinfo{author}{\bibfnamefont{M.~A.} \bibnamefont{Bates}},
  \bibinfo{journal}{Phys. Rev. E} \textbf{\bibinfo{volume}{74}},
  \bibinfo{pages}{061702} (\bibinfo{year}{2006}).

\bibitem[{\citenamefont{Matteis and Virga}(2005)}]{Matteis:pre:biax:tcp:2005}
\bibinfo{author}{\bibfnamefont{G.~D.} \bibnamefont{Matteis}} \bibnamefont{and}
  \bibinfo{author}{\bibfnamefont{E.~G.} \bibnamefont{Virga}},
  \bibinfo{journal}{Phys. Rev. E} \textbf{\bibinfo{volume}{71}},
  \bibinfo{pages}{061703} (\bibinfo{year}{2005}).

\end{thebibliography}


\end{document}